\documentclass[showpacs,prd,amsfonts,amsmath,twocolumn,aps,preprintnumbers,letterpaper]{revtex4-1}
\usepackage{amsmath,amssymb}
\usepackage{slashed}
\usepackage{graphicx}
\usepackage{epsfig}
\usepackage{braket}
\usepackage{bm}
\usepackage[top=1.0in,bottom=1.0in,left=1.0in,right=1.0in]{geometry}
\usepackage[caption=false]{subfig}
\usepackage[pdftex, pdfborder= 0 0 0, citecolor=blue, urlcolor=blue, linkcolor=blue, colorlinks=true, bookmarksopen=true]{hyperref}
\usepackage{tikz}
\usetikzlibrary{positioning,arrows,fit}
\usetikzlibrary{decorations.pathmorphing}
\usetikzlibrary{decorations.markings}
\tikzset{
fermion/.style={thick,draw=black, postaction={decorate},
    decoration={markings,mark=at position .5 with {\arrow[black]{triangle 45}}}},
pomeron/.style={thick,decorate,draw=black,
    decoration={coil,aspect=0}},
Odderon/.style={thick,decorate, draw=black,
    decoration={zigzag,aspect=0}},
photon/.style={thick,decorate, draw=black,
    decoration={coil,aspect=0}},
scalar/.style={thick,dashed, draw=black,
    },
 }

\newcommand{\be}{\begin{equation}}
\newcommand{\ee}{\end{equation}}
\newcommand{\bear}{\begin{eqnarray}}
\newcommand{\eear}{\end{eqnarray}}
\newcommand{\ba}{\begin{array}}
\newcommand{\ea}{\end{array}}

\def\be{\begin{eqnarray}}
\def\ee{\end{eqnarray}}
\def\bea{\be}
\def\eea{\ee}

\def\roughly#1{\mathrel{\raise.3ex\hbox{$#1$\kern-.75em%
\lower1ex\hbox{$\sim$}}}}

  \long\def\comment#1{ }

  \newcommand{\beq}{\begin{eqnarray}}
  \newcommand{\eeq}{\end{eqnarray}}
  
 \def\simge{\mathrel{%
   \rlap{\raise 0.511ex \hbox{$>$}}{\lower 0.511ex \hbox{$\sim$}}}}
\def\simle{\mathrel{
   \rlap{\raise 0.511ex \hbox{$<$}}{\lower 0.511ex \hbox{$\sim$}}}}

\newcommand*{\0}{\textcolor{red}}
\newcommand*{\1}{\textcolor{blue}}

\def\D0{D$\slashed{\mathrm{O}}$}
\begin{document}


\title{Threshold photoproduction of $\eta_c$ and $\eta_b$ using holographic QCD}

\author{Florian Hechenberger}
\email{florian.hechenberger@tuwien.ac.at}
\affiliation{Institut fur Theoretische Physik, Technische Universitat Wien, Wiedner Hauptstrasse 8-10, A-1040 Vienna, Austria
}

\author{Kiminad A. Mamo}
\email{kamamo@wm.edu}
\affiliation{
Physics Department, William \& Mary, Williamsburg, VA 23187, USA}

\author{Ismail Zahed}
\email{ismail.zahed@stonybrook.edu}
\affiliation{Center for Nuclear Theory, Department of Physics and Astronomy, Stony Brook University, Stony Brook, New York 11794-3800, USA}

\date{\today}

\begin{abstract}
We discuss the possibility that threshold photoproduction of $\eta_{c,b}$ may be sensitive to the pseudovector $1^{+-}$ glueball exchange. We use the holographic construction to identify the pseudovector glueball with the Kalb-Ramond field,  minimally coupled to bulk Dirac fermions. We derive  the holographic C-odd form factor and its respective charge radius. Using the pertinent Witten diagrams, we derive and analyze the differential photoproduction cross section for $\eta_{c,b}$ in the threshold regime, including the interference from 
the dual bulk photon exchange with manifest vector dominance. The possibility of measuring this process at current and future electron facilities is discussed.
\end{abstract}

\maketitle

%

\section{Introduction}
\label{Introduction}
At large center of mass energies, diffractive scattering of hadrons is dominated by Pomeron exchanges,  Reggeized even gluon exchanges with even C- and P-assignments. Initial perturbative QCD (pQCD) arguments suggest that odd gluon exchanges in the form of Odderon exchanges with odd C- and P-assignments 
are also possible~\cite{Braun:1998fs}
(and references therein). The signature of this exchange maybe observed in the difference 
between the diffractive $pp$ and $p\bar p$ cross sections, and the photoproduction or
electroproduction of heavy pseudoscalar mesons.

Recently, the TOTEM collaboration at the LHC, has reported a difference between their extrapolated $pp$ data at $\sqrt{s}=1.96\,{\rm GeV}$~\cite{TOTEM:2020zzr}, from the reported $p\bar p$ data by the \D0 collaboration 
at Fermilab, at the same center of mass energy. Their analysis suggests that the
difference is evidence for an Odderon. 
A number of recent analyses appear to also point in this direction~\cite{Royon:2023gkk} (and references therein).


At  weak coupling, the hard Pomeron is a Reggeized Balitskii-Fadin-Kuraev-Lipatov (BFKL) ladder which resums the rapidity ordered C-even collinear emissions. In the conformal limit, it is 
is identified  with the j-plane branch-points. By analogy,  the Odderon is a Reggeized BKP ladder which resums the C-odd collinear emissions~\cite{Bartels:1980pe,Kwiecinski:1980wb}. At strong coupling in dual gravity, the Pomeron is identified with a Reggeized spin-j graviton,
while the Odderon with a Reggeized spin-j Kalb-Ramond field~\cite{Brower:2008cy}. Further analyses of the gravity dual Odderon have been carried out in conformal geometries~\cite{Avsar:2009hc}, and more recently in confining geometries~\cite{Hechenberger:2023edv} with a detailed comparison to the recent TOTEM data.

The purpose of this work is to explore the possible contribution of the C-odd gluonic exchange, in the diffractive photoproduction of charmed and bottom pseudoscalars $\eta_{c,b}$ near threshold, using dual gravity. This approach was recently applied
to the description of photoproduction of charmonium near threshold at Jlab energies~\cite{Mamo:2019mka,Mamo:2022eui}, with relative success in extracting the mass and scalar radii of the
gluonic component of the nucleon~\cite{Duran:2022xag}.

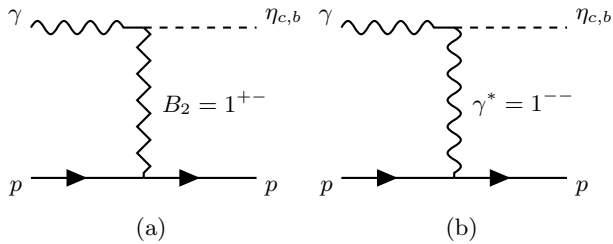
\begin{figure}[!htb]
    \begin{minipage}{.25\textwidth}
       \raggedleft
        \begin{tikzpicture}[node distance=1cm and 1.5cm]
            \coordinate[label={[yshift=4pt]left:$\gamma$}] (e1);
            \coordinate[right=of e1] (aux1);
            \coordinate[right=of aux1,label={[yshift=4pt]right:$\eta_{c,b}$}] (e2);
            \coordinate[below=2cm of aux1] (aux2);
            \coordinate[left=of aux2,label={[yshift=-4pt]left:$p$}] (e3);
            \coordinate[right=of aux2,label={[yshift=-4pt]right:$p$}] (e4);
            
            \draw[photon] (e1) -- (aux1);
            \draw[scalar] (aux1) -- (e2);
            \draw[fermion] (e3) -- (aux2);
            \draw[fermion] (aux2) -- (e4);
            \draw[Odderon] (aux1) -- node[label={right:$B_2=1^{+-}$}] {} (aux2);
        \end{tikzpicture}
        \centering (a)
        \label{fig:OdderonEx}
    \end{minipage}%
    \begin{minipage}{0.25\textwidth}
    \raggedright
        \begin{tikzpicture}[node distance=1cm and 1.5cm]
            \coordinate[label={[yshift=4pt]left:$\gamma$}] (e1);
            \coordinate[right=of e1] (aux1);
            \coordinate[right=of aux1,label={[yshift=4pt]right:$\eta_{c,b}$}] (e2);
            \coordinate[below=2cm of aux1] (aux2);
            \coordinate[left=of aux2,label={[yshift=-4pt]left:$p$}] (e3);
            \coordinate[right=of aux2,label={[yshift=-4pt]right:$p$}] (e4);
            
            \draw[photon] (e1) -- (aux1);
            \draw[scalar] (aux1) -- (e2);
            \draw[fermion] (e3) -- (aux2);
            \draw[fermion] (aux2) -- (e4);
            \draw[photon] (aux1) -- node[label={right:$\gamma^*=1^{--}$}] {} (aux2);
        \end{tikzpicture}
        \centering (b)
        \label{fig:photonEx}  
   \end{minipage}
   \caption{threshold photoproduction of  $\eta_c$ through (a) Kalb-Ramond exchange and (b) P-wave photon exchange}
   \label{fig:PO}
\end{figure}

In dual gravity, threshold charmoium photoproduction is dominated by the C-even 
and $2^{++}$ glueball exchange, with some admixture of $0^{++}$ glueball exchange 
for large skewness. Similarly, we expect that threshold photoproduction of charmed 
pseudoscalars to be dominated by C-odd $1^{+-}$ glueball exchanges, modulo 
the photon Primakoff exchange as depicted in Fig.~\ref{fig:PO}. This process
provides for a possible measure of the C-odd gluon charge radius.

The organization of the paper is as follows: in section~\ref{SEC2} we outline
the dual bulk action for the the photoproduction of heavy $eta$ pseudoscalars. Following the initial suggestion in~~\cite{Brower:2008cy}, the C-odd gluon exchange is identified with the Kalb-Ramond
2-form in the bulk, with coupling to the photon and pseudoscaars governed by the Chern-Simons term. In section~\ref{SEC3} the dual photoproduction amplitudes are evaluated using the leading Witten diagrams. The
C-odd bulk form factors are explicitly derived, and the corresponding charge radii
derived. In section~\ref{SEC4} we detail the differential cross section for photoproduction in the treshod region. Detail numerical results are presented at currently electron machines. Our conclusions are in section~\ref{SEC5}. We have added a number of Appendices to detail some of the derivations in the main text.

\section{Bulk action}
\label{SEC2}
We recall that the $1^{+-}$ Kalb-Ramond as a 2-form field, couples to the light flavor brane through the Chern-Simons term.
For instance, for flavor D8 probe branes, 
\begin{widetext}
\bea
S_{CS}=T_8\int {\rm Tr}\bigg(e^{\mathbb F}\wedge \sum_jC_{2j+1}\bigg)\rightarrow 
\tilde{T}_8\int d^5x \epsilon^{MNOPQ}{\rm Tr}\big(A_M F_{NO}B_{PQ})
\label{eq:CS}
\eea
with the 2-form $\mathbb F=2\pi\alpha^\prime F+B$,
 the sum of  $F=dA-iA^2$ the flavor 2-form  and the $1^{+-}$ Kalb-Ramond 2-form $B$.
 The tension of the D8 brane is denoted by $T_8$. The light $\eta$ field is usually identified 
 with the singlet part of $A_z=\frac 1{\sqrt N_f} \eta$, with $N_f=1+2$ 
 (for a single heavy and two light flavor branes),  while the $U(1)$ gauge field with the space-time parts of $F$. We
 will assume that an analogous coupling carries to the heavy $\eta_{c,b}$. The bulk action relevant to the photoproduction of $\eta_{c,b}$ reads
\bea
\label{XACTIONB2}
S_B=\int d^5x\,\sqrt{g} &\bigg(&-\frac 1{12\tilde{g}_5^2}e^{-2\phi}H^{MNO}H_{MNO}\nonumber\\
&&+\frac 12 g_{CS}\, {\rm Tr}\,\epsilon^{MNOP}A_z F_{MN}B_{OP}+e^{-\phi}g_{B\psi} \sum_{1,2} (\pm)\overline\Psi_{1,2}  e^M_Ae^N_B\sigma^{AB} \Psi_{1,2}\,B_{MN}\bigg)\,,
\eea
for the Kalb-Ramond field $B_2$ and
\bea
\label{XACTIONX}
S_A=\int d^5x\, \sqrt{g}e^{-\phi}&\bigg(&-\frac 1{4g_5^2}F^{MN}F_{MN}+\sum_{1,2}\frac{i}{2g_5^2}\overline{\Psi}_{1,2}e^M_A\Gamma^A\Psi_{1,2} A_M\nonumber\\
&&+\frac 12 g_{CS}\, {\rm Tr}\,\epsilon^{MNOP}F_{MN}F_{OP}+e^{-\phi}\eta_P\sum_{1,2}(\pm)\overline\Psi_{1,2}  e^M_Ae^N_B\sigma^{AB} \Psi_{1,2}\,F_{MN}\bigg),\nonumber\\
\eea
\end{widetext}
for the vector mesons, which will supply the relevant photon couplings through vector-meson-dominance (VMD). The coupling $g_{CS}$ is uniquely determined by the 5D Chern-Simons term to be $g_{CS}=\frac{N_c}{24\pi^2}$. The 5D Newton constant is given by $\tilde{g_5}^2=2\kappa^2=16\pi G_N=8\pi^2/N_c^2$ and $\eta_P$ is the Pauli parameter, which will be fixed by matching the Pauli form factor to its experimental value. In the above equations the flavor trace will pick up the relevant charges for charmonia $e_c=2/3~e$ or bottomonia $e_b=-1/3~e$.
Here $H_3=dB$ is the 3-form field strength of the $1^{+-}$ Kalb-Ramond field. The background is given by the AdS metric
\bea
ds^2=
\frac{L^2}{z^2}(dx^2+dz^2),
\label{eq:AdSmetric}
\eea
with non-constant dilaton $\phi=\kappa^2z^2$. In the fermionic parts of the action we denote $\sigma^{AB}=\frac{i}{2}\left[\Gamma^A,\Gamma^B\right]$, with the gamma matrices given by $ \Gamma^A=(\gamma^\mu,-i\gamma^5)$ and obeying the Clifford algebra $\left\{\Gamma^A,\Gamma^B\right\}=2\eta^{AB}$. The tetrads following from \eqref{eq:AdSmetric} are given by $e^M_A=z\delta^M_A$.
The positive and negative parity Dirac spinors follow from the mixed representation of (\ref{SolutionFermions}) in Appendix \ref{app:BulkFields}, to which we refer the interested reader for further details.
The axial gauge field $V_M=(0,V_\mu)$ is the projected spin-1  axial-field 
$$B_{\mu\nu}= \frac 1{\sqrt{-\partial^2}}\epsilon_{\mu\nu\rho\sigma}\partial^\rho V^\sigma$$ with the physical polarization 
$\langle 0|V_\mu|V;P\rangle=\epsilon_\mu(P)$. The projection yields the 3 physical degrees of freedom out of the 6 gauge degrees of freedom in $B$, and guarantees the
correct normalization for the ensuing kinetic term. We have included the sole coupling to a bulk Dirac fermion through its magnetic moment, as suggested by supergravity (SUGRA). In Appendix~\ref{SSTYPEII} we give the triple couplings $1^{\pm -}\eta\gamma$ in the Sakai-Sugimoto model for comparison. The dual field with boundary spin values
$$\tilde{B}^{\mu\nu}=\frac{1}{2}\epsilon^{\mu\nu\alpha\beta}B_{\alpha\beta}=C^{\mu\nu}$$ carries $1^{--}$ assignment. Following~\cite{Brower:2008cy},
we make the boundary identifications with scalar and pseudo-scalar gluonic operators with mass dimension $\Delta=6$
\bea
\label{BOUNDARYOP}
B^{\mu\nu}&\rightarrow& d^{abc}G^{a\alpha\beta} G^b_{\alpha\beta}G^{c\mu\nu}\nonumber\\
{\tilde{B}}^{\mu\nu}&\rightarrow& d^{abc}G^{a\alpha\beta} G^b_{\alpha\beta}{\tilde{G}}^{c\mu\nu}
\eea
which we interpret as C-odd twist-5 operators  on the light front. In contrast, 
we note that in the context of pQCD, factorization arguments show that the leading 
contribution to the photoproduction of $\eta_c$, in the large skewness limit,  is a C-odd local twist-3 operator~\cite{Ma:2003py}
$$d^{abc}G^{a\alpha +} G^{b+}_{\alpha}G^{c+\nu}$$



\section{$\eta_c$ and $\eta_b$ photoproduction in dual gravity}
\label{SEC3}
In dual gravity, threshold photoproduction of  $\eta_{c,b}$ by exchange of a Kalb-Ramond field is illustrated by the Witten diagram in Fig.~\ref{fig:hPO}a. This process
is very similar to the photoproduction of charmonium, with similar kinematics given
the $\eta_c$ mass of 2.984 GeV, and  the $J/\Psi$  mass of 3.097 GeV.  The  essential differences stem from their quantum numbers: P-odd versus P-even couplings, with the former expected to be more suppressed. In light of this, and motivated by previous analyses \cite{Czyzewski:1996bv,Dumitru:2019qec,Bartels:2001hw,Jia:2022oyl}, we have also included the
tree level Witten diagram contribution stemming from the exchange of a bulk photon 
in Fig.~\ref{fig:hPO}b.

 \subsection{Dual photoproduction amplitude}
\begin{figure*}[!htb]
    \begin{minipage}{.45\textwidth}
       \centering
        \begin{tikzpicture}[node distance=1cm and 1.5cm]
            \coordinate[label={[yshift=4pt]left:$\mathcal{V}(q_1;z')$}] (e1);
            \coordinate[below right=of e1] (aux1);
            \coordinate[above right=of aux1,label={[yshift=4pt]right:$\eta_{c,b}(m_{\eta_{c,b}};z')$}] (e2);
            \coordinate[below=1.25cm of aux1] (aux2);
            \coordinate[below left=of aux2,label={[yshift=-4pt]left:$\Psi(p_1;z)$}] (e3);
            \coordinate[below right=of aux2,label={[yshift=-4pt]right:$\overline{\Psi}(p_2;z)$}] (e4);
            \draw[photon] (e1) -- (aux1);
            \draw[scalar] (aux1) -- (e2);
            \draw[fermion] (e3) -- (aux2);
            \draw[fermion] (aux2) -- (e4);
            \draw[Odderon] (aux1) -- node[label={right:$G_1^{\mathbb O}(s,t,z,z')$}] {} (aux2);
            \node[draw, thick,circle,fit=(e1) (e4),inner sep=.5\pgflinewidth] {};
        \end{tikzpicture}
        \center{(a)}
        \label{fig:hOdderonEx}
    \end{minipage}
    \begin{minipage}{0.45\textwidth}
    \centering
            \begin{tikzpicture}[node distance=1cm and 1.5cm]
            \coordinate[label={[yshift=4pt]left:$\mathcal{V}(q_1;z')$}] (e1);
            \coordinate[below right=of e1] (aux1);
            \coordinate[above right=of aux1,label={[yshift=4pt]right:$\eta_{c,b}(m_{\eta_{c,b}};z')$}] (e2);
            \coordinate[below=1.25cm of aux1] (aux2);
            \coordinate[below left=of aux2,label={[yshift=-4pt]left:$\Psi(p_1;z)$}] (e3);
            \coordinate[below right=of aux2,label={[yshift=-4pt]right:$\overline{\Psi}(p_2;z)$}] (e4);
            \draw[photon] (e1) -- (aux1);
            \draw[scalar] (aux1) -- (e2);
            \draw[fermion] (e3) -- (aux2);
            \draw[fermion] (aux2) -- (e4);
            \draw[photon] (aux1) -- node[label={right:$G_1^\gamma(s,t,z,z')$}] {} (aux2);
            \node[draw, thick,circle,fit=(e1) (e4),inner sep=.5\pgflinewidth] {};
        \end{tikzpicture}
        \center{(b)}
        \label{fig:hPhotonEx}  
   \end{minipage}
   \caption{Witten diagrams for threshold production of $\eta_{c,b}$ through (a) Odderon and (b) photon exchange}
    \label{fig:hPO}
\end{figure*}
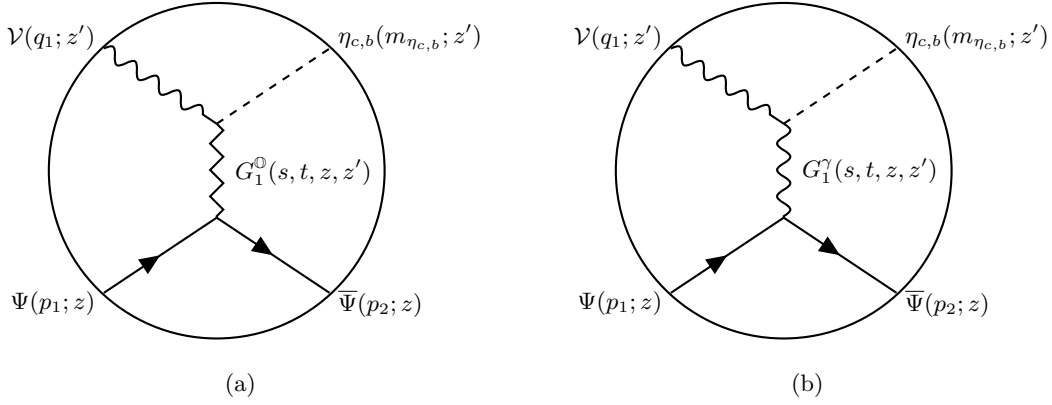 
Using (\ref{XACTIONX}), the  Witten diagram in Fig.~\ref{fig:hPO}a yields the 
photoproduction amplitude for $\eta_{c}$ as 
\begin{widetext}
\bea
    i{\cal A}(s,t)_{\gamma p\to \eta p}&=&\sum_n i\tilde{{\cal A}}_{\gamma p\to \eta p}(m_n,s,t)\\
    i\tilde{{\cal A}}_{\gamma p\to \eta p}(m_n,s,t)&=&(-i)V^\mu_{\mathbb{O}\gamma\eta}(q_1,q_2,k,m_n)\times\tilde P_{\mu\nu}(m_n^{\mathbb{O}},\Delta)\times(-i)V^\nu_{\mathbb{O}\overline{\Psi}\Psi}(p_1,p_2,k,m_n)\nonumber\\
    &&+(-i)V^\mu_{\gamma\gamma^*\eta}(q_1,q_2,k, m_n)\times\tilde P_{\mu\nu}( m_n^\gamma,\Delta)\times(-i)V^\nu_{\gamma*\overline{\Psi}\Psi}(p_1,p_2,k,m_n),\nonumber
    \label{eq:amp}
\eea   
with the bulk vertices
\bea
     V^\mu_{\mathbb{O}\gamma\eta}(q_1,q_2,k)&=&
     g_{CS}\frac{e_{c,b}}{2\sqrt{N_f K^2}}\int dz \phi(z)J_b(m_n,z)\epsilon^{\mu\nu\rho\sigma}k_\nu F_{\rho\sigma}\nonumber\\ 
    V^{\sigma}_{\mathbb{O}\overline{\Psi}\Psi}(p_1,p_2,k)&=&
    \frac{1}{2\sqrt{K^2}}\int dz\sqrt{g}e^{-\phi}\sum_{1,2}(\pm) \overline\Psi_{1,2} (p_2,z)\sigma^{\mu\nu}\Psi_{1,2}(p_1,z) J_b(m_n,z)\epsilon_{\mu\nu\rho\lambda}k^\rho\eta^{\lambda\sigma},
    \label{eq:normOddVertex}
\eea
where the field strength is now to be understood as $F_{\rho\sigma}=iq_\rho\epsilon_\sigma(q)-iq_\sigma\epsilon_\rho(q)$ with $\epsilon_\mu(q)$ is the polarization of the external photon with momentum $q$. The massive spin-1 propagator in the mode-sum representation is given by
\bea
G_1(m_n,t,z,z')_{\mu\nu}&=&J_b(m_n,z)\tilde{P}_{\mu\nu}J_b(m_n,z'),\nonumber\\
\tilde P_{\mu\nu}(m_n,k)&=&\frac{-i}{k^2-m_n^2}P_{\mu\nu}(k)\nonumber\\
P_{\mu\nu}(k)&=&\eta^{\mu\nu}-\frac{k^\mu k^\nu}{k^2}.
\eea
Similarly, we obtain for the photon vertices in the bulk
\bea
    V^\mu_{\gamma^\star\gamma\eta}(q_1,q_2,K)&=&\frac{e_{c,b}^2}{2\sqrt{N_f}}\int dz\phi(z)\times\frac{z^2}{2}\epsilon^{\mu\nu\rho\sigma}F_{\mu\nu}F_{\rho\sigma}\\
    V^{\nu(1)}_{\gamma^\star\overline{\Psi}\Psi}(p_1,p_2,K)&=&\frac{e_{c,b}}{2g_5^2}\int dz\sqrt{g}e^{-\phi}\sum_{1,2} \overline\Psi_\pm (p_2,z)\gamma^\nu\Psi_\pm(p_1,z)J(m_n,z)\nonumber\\
    V^{\nu(2)}_{\gamma^\star\overline{\Psi}\Psi}(p_1,p_2,K)&=&\eta_P\frac{e}{2}\int dz\sqrt{g}e^{-\phi}\sum_{1,2} (\pm)\overline\Psi_\pm (p_2,z)\sigma^{\mu\nu}\Psi_\pm (p_1,z)K_\mu J(m_n,z)\nonumber\\
    V^{\nu(3)}_{\gamma^\star\overline{\Psi}\Psi}(p_1,p_2,K)&=&\eta_P\frac{e}{2}\int dz\sqrt{g}e^{-\phi}\sum_{1,2} \overline\Psi_\pm (p_2,z)\gamma^\nu i\gamma^5\Psi_\pm(p_1,z)\partial_z J(m_n,z)\nonumber\\
    \label{eq:nonnormGaVertex}
\eea
The bulk coupling to $\eta_{c,b}$ is also governed by the Chern-Simons term in \eqref{eq:CS} via the substitution $B_{MN}\to F_{MN}$. Note that there are no metric and dilaton factors in the coupling of $B_2$ and $A_\mu$ to $\gamma-\eta_{c,b}$ in \eqref{eq:normOddVertex} since this interaction is purely governed by the Chern-Simons term in \eqref{eq:CS}. The baryon couplings on the other hand are governed by the Dirac-Born-Infeld (DBI) part of the action and only receive $1/N_c$ corrections from the Chern-Simons term. 
The photon couplings follow analogously with $\kappa_b$ replaced by $\kappa$.

For $z'\to 0$ and $t=-K^2$ we can use \eqref{eq:b2bSpacelike} in \eqref{eq:amp} to get
\bea
    i{\cal A}(s,t)_{\gamma p\to \eta p}&=& i\tilde{{\cal A}}_{\gamma p\to \eta p}(s,t)\\
    i\tilde{{\cal A}}_{\gamma p\to \eta p}(s,t)&=&(-i)\mathcal{V}^\mu_{\mathbb{O}\gamma\eta}(q_1,q_2,k)\times P_{\mu\nu}(\Delta)\times(-i)\mathcal{V}^\nu_{\mathbb{O}\overline{\Psi}\Psi}(p_1,p_2,k)\\
    &&+(-i)\mathcal{V}^\mu_{\gamma\gamma^*\eta}(q_1,q_2,k)\times P_{\mu\nu}(\Delta)\times(-i)\mathcal{V}^\nu_{\gamma*\overline{\Psi}\Psi}(p_1,p_2,k)
    \label{eq:ampK}
\eea
With the normalizable modes in \eqref{eq:normOddVertex} now substituted with their non-normalizable counterparts ${\cal V}(Q,z)$. In the space-like region, we  obtain for the Kalb-Ramond amplitude 
\bea
    \mathcal{V}^\mu_{\mathbb{O}\gamma\eta}(q_1,q_2,K)&=&g_{CS}\frac{e_{c,b}}{2\sqrt{N_f}}\int dz\varphi(z)\times\frac{z^2}{2}\epsilon^{\mu\nu\rho\sigma} k_\nu F_{\rho\sigma}\nonumber\\ 
    \mathcal{V}^{\sigma}_{\mathbb{O}\overline{\Psi}\Psi}(p_1,p_2,K)&=&\frac{g_{B\Psi}}{2\sqrt{K^2}}\int dz\sqrt{g}e^{-\phi}\sum_{1,2}(\pm) \overline\Psi_{1,2} (p_2,z)\sigma^{\mu\nu}\Psi_{1,2}(p_1,z) \mathcal{V}_b(K,z)\epsilon_{\mu\nu\rho\lambda}
K^\rho\eta^{\lambda\sigma}\nonumber\\
    &=&\frac{g_{B\Psi}}{2\sqrt{K^2}}\int dz\sqrt{g}e^{-\phi}2\psi_L(z)\psi_R(z)\mathcal{V}_b(K,z)\overline{u}(p_2)\gamma^5\sigma_{\rho\lambda}u(p_1)
K^\rho\eta^{\lambda\sigma}.\nonumber\\
    \label{eq:nonnormOddVertex}
\eea
and for the photon amplitude analogously 
\bea
    \mathcal{V}^\mu_{\gamma^\star\gamma\eta}(q_1,q_2,K)&=&\frac{e_{c,b}^2}{2\sqrt{N_f}}\int dz\varphi(z)\times\frac{z^2}{2}\epsilon^{\mu\nu\rho\sigma}\eta F_{\mu\nu}F_{\rho\sigma}\\
    \mathcal{V}^{\nu(1)}_{\gamma^\star\overline{\Psi}\Psi}(p_1,p_2,K)&=&\frac{e_{c,b}}{2g_5^2}\int dz\sqrt{g}e^{-\phi}\sum_{1,2} \overline\Psi_\pm (p_2,z)\gamma^\nu\Psi_\pm(p_1,z){\cal V}(K,z)\nonumber\\
    &=&\frac{e}{2g_5^2}\int dz\sqrt{g}e^{-\phi}z\left(\psi_R^2(z)+\psi_L^2(z)\right){\cal V}(K,z)\overline u(p_2)\gamma^\nu u(p_1)\nonumber\\
    \mathcal{V}^{\nu(2)}_{\gamma^\star\overline{\Psi}\Psi}(p_1,p_2,K)&=&\eta_P\frac{e}{2}\int dz\sqrt{g}e^{-\phi}\sum_{1,2} (\pm)\overline\Psi_\pm (p_2,z)\sigma^{\mu\nu}\Psi_\pm (p_1,z)K_\mu{\cal V}(K,z)\nonumber\\
    &=&\eta_P\frac{e}{2}\int dz\sqrt{g}e^{-\phi}\left(2\psi_L(z)\psi_R(z)\right){\cal V}(K,z)\overline u(p_2)\sigma^{\mu\nu}u(p_1)K_\mu\nonumber\\
    \mathcal{V}^{\nu(3)}_{\gamma^\star\overline{\Psi}\Psi}(p_1,p_2,K)&=&\eta_P\frac{e}{2}\int dz\sqrt{g}e^{-\phi}\sum_{1,2} \overline\Psi_\pm (p_2,z)\gamma^\nu i\gamma^5\Psi_\pm(p_1,z)\partial_z{\cal V}(K,z)\nonumber\\
    &=&\eta_P\frac{e}{2}\int dz\sqrt{g}e^{-\phi}z\left(\psi_L^2(z)-\psi_R^2(z)\right){\cal V}(K,z)\overline u(p_2)\gamma^\nu u(p_1).\nonumber\\
\eea

\subsection{Dual form factors}
The form factors are extracted from the 3-point functions with pertinent Lehmann–Symanzik–Zimmermann (LSZ) reduction. For example, the Dirac form factor resulting from the current associated with the covariant derivative receivs contributions from
\bea
W^\mu(K^2)^{EM}_{Dirac}=\overline{u}(p_2)\gamma^\mu u(p_1)\times e_N\times C_1(K)\equiv\frac{1}{F_N(p_2)F_N(p_1)}\frac{\delta S^{EM}_{Dirac}}{\delta \epsilon_\mu}\nonumber\\
\eea
with $e_N=e$ for the proton and $e_N=0$ for the neutron, $F_N(p)=\bra{0}\mathcal{O}_N(0)\ket{N(p)}$ the nucleon source constant and
\end{widetext}
\be
C_1(K)&=&\frac{1}{2}\int e^{-\kappa^2z^2}z^{3-2\tau}(\tilde{\psi}_L^2+\tilde{\psi}_R^2)\mathcal{V}(Q,z)\nonumber\\
&=&\frac{(a_K+2\tau)\Gamma(a_K+1)\Gamma(\tau)}{2\Gamma(a_K+\tau+1)}.\nonumber\\
\ee
Similar relations hold for the other 3-point amplitudes. The electromagnetic Dirac and Pauli form factors are thus given by
\bea
F_1(Q)&=&C_1(K)+\eta_P C_2(K)\nonumber\\
F_2(Q)&=&\eta_P C_3(K)
\eea
where
\bea
C_2(K)&=&\frac{1}{2}\int e^{-\kappa^2z^2}z^{3-2\tau}(\tilde{\psi}_L^2-\tilde{\psi}_R^2)z\partial_z\mathcal{V}(K,z)\nonumber\\
&=&\frac{a_K(a_K(\tau-1)-1)\Gamma(a_K+1)\Gamma(\tau)}{\Gamma(a_K+\tau+2)}\nonumber\\
C_3(K)&=&2M_N\int e^{-\kappa^2z^2}z^{3-2\tau}\tilde{\psi}_L\tilde{\psi}_R z\mathcal{V}(Q,z)\nonumber\\
&=&\frac{4(\tau-1)\tau\Gamma(a_K+1)\Gamma(\tau)}{\Gamma(a_K+\tau+1)}
\eea
which follow from
\be
&&\big\langle N(p_2)|J^\mu_{EM}(0)|N(p_1)\big\rangle=\nonumber\\
&&\overline{u}(p_2)\left(F_1(K)\gamma^\mu+F_2(K)\frac{i\sigma^{\mu\nu}}{2M_N}k_\nu\right)u(p_1),\nonumber\\
\ee  
as previously obtained in \cite{Mamo:2021jhj}. Note the appearance of an additional contribution to $F_1(Q)$ from the 5D Pauli term $\sigma^{\mu z}$. The proton electromagnetic form factor normalizations are fixed by the charge
$F_1(0)=1$ (Dirac)  and magnetic moment given in units of the nuclear magneton
$$F_2(0)=(\mu_p-1)=1.7928\qquad {\rm  (Pauli)},$$
where we used $\mu_p/\mu_N=2.7928$. This fixes $\eta_P=1.7928/C_3(0)=1.7928/4(\tau-1)$. Similarly, the C-odd Kalb-Ramond or Odderon form factor is given by
\bea
F_b(K)&=&
\int dzz^{-2\tau+3}e^{-\phi}2M_N\tilde{\psi}_R\tilde{\psi}_L\mathcal{V}_b(K,z)\nonumber\\
&=&16 (\tau -1) \Gamma (\tau +1) \Gamma \left(a_K+1\right)\nonumber\\
&&\times_2\tilde{F}_1\left(\tau +1,a_K+1;\tau +a_K+1;-3\right)\nonumber\\
\eea
where we pulled out a facotr of $2M_N$ to highlight the similarity with the electromagnetic Pauli form factor and $_p\tilde{F}_q$ is the regularized hypergeometric function. The ensuing C-odd squared charge radius is
\bea
\langle r^2\rangle=-6\bigg(\frac{d{\rm log}F_b(K)}{dK^2}\bigg)_{K^2=0}\label{eq:CoddChargeRadius}
\eea
The C-odd form factor normalizations are fixed by the nucleon tensor charge (axial-Pauli) and  the nucleon intrinsic spin (axial-Dirac).
More specifically, the nucleon tensor charge is defined by the matrix element
\bea
\label{TENSORCHARGE}
\langle P\, S|\bar\psi i\sigma^{\mu\nu}\gamma^5\psi|P\,S\rangle= 
2\delta q(P^\mu S^\nu-P^\nu S^\mu)\nonumber\\
\eea
At a resolution of the order of the nucleon mass, lattice evaluation gives~\cite{Aoki:1996pi} 
\bea
\delta q=\delta u+\delta d\approx 0.839-0.231=0.608
\eea
The intrinsic spin of the nucleon is mostly due to the mixing with the gluons from
the $U(1)_A$ anomaly
\bea
\langle P\, S|\bar\psi i\gamma^\mu\gamma^5\psi|P\,S\rangle= 2M_N\Sigma(0)S^\mu.
\eea
The estimation from the QCD instanton vacuum gives
$\Sigma(0)=0.3$ at a resolution of about the nucleon mass~\cite{Zahed:2022wae}, while lattice simulations give $\Sigma(0)=0.4$, at a resolution of about twice the nucleon mass~\cite{Alexandrou:2017oeh}. Therefore, at the nucleon mass resolution, we set  
\be
F_b(0)=0.608
\label{eq:CoddNorm}
\ee
Using \eqref{eq:CoddChargeRadius} we readily obtain the charge radius
\bea
\langle r^2\rangle&=&\frac{3}{2 \kappa _b^2} \bigg(\gamma_E -4 \Gamma (\tau +1)\nonumber\\
&&\times\bigg(\ _2F_1^{(0,0,1,0)}(1,\tau +1,\tau +1,-3)\nonumber\\
&&+\ _2F_1^{(1,0,0,0)}(1,\tau +1,\tau
   +1,-3)\bigg)\bigg)\nonumber\\
\eea   
where $\gamma_E$ is the Euler-Mascheroni constant. For $\tau=3$ and $\kappa_\gamma=0.3875\text{ GeV}$ we obtain
\bea
\sqrt{\langle r^2\rangle} &=&2.733\text{ GeV}^{-1}=0.540\text{ fm},\nonumber\\
\eea
For comparison, we note that the Odderon-nucleon  coupling as a C-odd and un-Reggeized 3-gluon exchange in~\cite{Czyzewski:1996bv}, is assumed monopole-like  with unit normalization. 
Also in the eikonal dipole analysis at low-x~\cite{Dumitru:2019qec}, the Odderon-nucleon form factor  is argued to be fixed by the leading twist quark generalized parton distribution (GPD), with a normalization to 1. 
In contrast, the Reggeized BKP Odderon-nucleon form factor
in~\cite{Bartels:2001hw} is relatively large, with even 
a rapid sign change at the origin. 

\begin{figure}
    \centering
    \includegraphics[height=5cm,width=8cm]{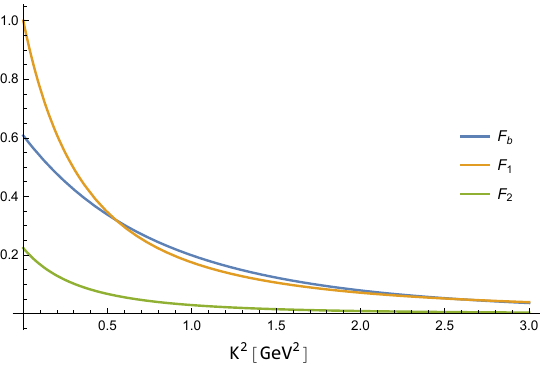}
    \caption{C-even and C-odd nucleon form factors in the approximation $\kappa_N=\kappa_\gamma$ with the normalization fixed by the charge, magnetic moment  and \eqref{eq:CoddNorm}.}
    \label{fig:formFactors}
\end{figure}

The form factors are displayed in Fig.~\ref{fig:formFactors} with $\phi=\kappa_N^2z^2=\kappa_\gamma^2z^2=\kappa^2z^2$ for the open string sector and $\phi=\kappa_b^2z^2=4\kappa^2z^2$ for the closed string sector.  We fix $\kappa$ by the $\rho$ meson pole in the (time-like) photon bulk-to-boundary propagator, as is required by VMD, giving
\be
(\kappa_{b}, \kappa_{\gamma}, \kappa_{N})=(0.775, 0.3875,0.3875)\,{\rm GeV}.\nonumber\\
\ee
For moderate $K^2$ the dominant contribution on the light front stems from the $F_1$ contribution of the photon, in analogy to, but not as pronounced as, the Primakoff effect. This is due to the absence of the photon pole in VMD. At larger $K^2$ the C-odd contribution will dominate the differential cross section due to the kinematical nature of the coupling in \eqref{XACTIONB2}, in agreement  with pQCD calculations \cite{Jia:2022oyl,Dumitru:2019qec}. 

\subsection{Threshold vertices}
At threshold, the Odderon-$\eta_{c,b}\gamma$ vertices in the space-like region is given by
\bea
\mathcal{V}_{B\eta\gamma}(K)&=&\frac{e_{c,b}}{2\sqrt{N_f}}\int dz\varphi(z){\cal V}_b(K,z).\nonumber\\
\eea
The pertinent LSZ reduction for the production of $\eta_{c,b}$ at the boundary results in a substitution rule for the bulk-to-boundary propagator
\bea
\varphi(q,z)\to\phi_n(z)&=&g_5c_n\kappa z L_n(\kappa^2z^2)\nonumber\\
&=&-\frac{f_n}{m_n}\times 4g_5(n+1)\kappa z L_n(\kappa^2z^2),\nonumber\\
\eea
hence reducing to a simple vertex factor
\be
\mathcal{V}_{B\eta\gamma}(K)&\approx& \frac{e_{c,b}}{2\sqrt{N_f}}\int dz \phi_n(z)\times \frac{z^2}{2}\nonumber\\
&\equiv& e_{c,b}\left(\frac{f_{\eta_{c,b}}}{m_{\eta_{c,b}}}\right)\mathbb{V}_{B\eta\gamma}.\nonumber\\
\label{eq:OdderonEtaGammaRed}
\ee
The absence of the dilaton/metric in the Chern-Simons term
implies that (\ref{eq:OdderonEtaGammaRed}) is divergent in the IR.
Note that this is not the case if we were to use a slab geometry
with a hard wall.  With this in mind, 
to obtain an estimate for the coupling ${\mathbb{V}_{B\eta\gamma}}$, we use the simple hard-wall cutoff obtained from \eqref{eq:HWmassSpectrum1} to obtain ${\mathbb{V}_{B\eta\gamma}}=-\frac{g_5}{\sqrt{N_f}}\kappa_\gamma z_0^4/4$.

The vertex containing a single virtual and one real photon is given by
\be
\mathcal{V}_{\eta\gamma\gamma^*}(K)=\frac{e_{c,b}^2}{2\sqrt{N_f}}\int dz \varphi_n(z){\cal V}(K,z)
\ee
As is the case for the Kalb-Ramond field, LSZ reduction picks out the pertinent normalizable mode and we obtain the same vertex as in \eqref{eq:OdderonEtaGammaRed}
\be
\mathcal{V}_{\eta\gamma\gamma^*}(K)&\approx& \frac{e_{c,b}^2}{2\sqrt{N_f}}\int dz \phi_n(z)\times \frac{z^2}{2}\nonumber\\
&\equiv& e_{c,b}^2\left(\frac{f_{\eta_{c,b}}}{m_{\eta_{c,b}}}\right)\mathbb{V}_{\gamma\gamma\eta}.\nonumber\\
\ee
with $\mathbb{V}_{\gamma\gamma\eta}=-\frac{g_5}{\sqrt{N_f}}\kappa_\gamma z_0^4/4$. For the numerical analysis we fix the decay constant $f_c$ by the leading order decay rate from pQCD~\cite{Pham:2007xx}
\be
\Gamma_{\eta\to\gamma\gamma}=4\pi Q_c^4\alpha^2\frac{f_\eta^2}{M_\eta}
\ee
with $Q_c$ the charm quark charge. From the experimental value $\Gamma_{\eta_c\to\gamma\gamma}=5.376\times 10^{-6}$ GeV~\cite{Workman:2022ynf} we obtain
\be
f_{\eta_c}=0.327\ \rm GeV,
\ee
where we used $M_{\eta_c}=2.9839$ GeV. The value for $\Gamma_{\eta_b\to\gamma\gamma}$ is not reported. However, from heavy quark symmetry, it follows that
\bea
\frac{f_{\eta_b}}{f_{\eta_c}}=\sqrt{\frac {M_{\eta_c}}{M_{\eta_b}}},
\eea
which amounts to
\be
f_{\eta_b}=0.184\ \rm GeV,
\ee
with $M_{\eta_b}=9.3897$ GeV.

\section{Differential cross section}
\label{SEC4}
The differential cross section is obtained by averaging over the initial state spins and polarizations and by summing over the final state spins
\be
\frac{d\sigma}{dt}=\frac{1}{16\pi(s-M_N^2)^2}\frac{1}{2}\sum_{\rm pol}\frac{1}{2}\sum_{\rm spin}\left|{\cal A}(s,t)_{\gamma p\to \eta p}\right|^2.\nonumber\\
\ee
The cross sections are then obtain from
\be
\label{SIGMATX}
\sigma(s)=\int_{-t_{\rm min}}^{-t_{\rm max}}\frac{d\sigma}{dt},
\ee
with $t_{min/max}$ fixed by the kinematics of the process, which are detailed in Appendix \ref{app:kin}.

Carrying out the polarization and fermion spin sums we arrive at
\begin{widetext}
\bea
\label{DSIGMAT}
\frac{d\sigma}{dt}
=\frac{2e^2e_{c,b}^2g_{CS}^2}{16\pi(s-M_N^2)^2}\times\left(\frac{f_X}{M_X}\right)^2\times\,
\, \bigg(F_{\mathbb O}(s,t,M_N,M_X)+F_{\gamma}(s,t,M_N,M_X)+F_{{\mathbb O}\gamma}(s,t,M_N,M_X)\bigg),\nonumber\\
\eea
with respectively, the C-even $F_{\mathbb O}$, photon $F_{\gamma}$, and mixed
$F_{\mathbb O\gamma}$   contributions
\bea
F_{\mathbb O}(s,t,M_N,M_X)&=&-\frac{F_b(K)^2 g_{B \psi }^2 \mathbb{V}_{B \gamma  \eta }^2}{K^2
   M_N^2} \bigg(K^2 s (K^2+2 M_N^2+M_X^2)+M_N^2 K^2 (M_X^2-M_N^2)+M_X^4-K^2 s^2\bigg)\nonumber\\
 F_{\gamma}(s,t,M_N,M_X)&=&e^4e_X^2\mathbb{V}_{\gamma\gamma\eta}^2\nonumber\\
 &&\times\bigg(\frac{F_2(K)^2 K^2 (-K^2 s (K^2+2 M_N^2+M_X^2)+M_N^2 (K^2 (2 K^2+M_N^2)+3 K^2 M_X^2+M_X^4)+K^2 s^2)}{M_N^2}\nonumber\\
 &&+4 F_2(K) F_1(K) K^2
   (K^2+M_X^2)^2+2 F_1(K)^2 \bigg(K^6+2 K^4 (M_X^2-s)\nonumber\\
   &&+K^2 (-2 M_N^2 (M_X^2+2 s)+2 M_N^4-2 s M_X^2+M_X^4+2 s^2)-2 M_N^2 M_X^4\bigg)\bigg)\nonumber\\
F_{\mathbb O\gamma}(s,t,M_N,M_X)&=&0
\eea  
\end{widetext}
with zero mixing between the tensor contributions.

\subsection{Estimate of $g_{B\psi}$}
The  overall magnitude of the cross section (\ref{SIGMATX}) hinges considerably on the value of the bulk coupling $g_{B\psi}$. For an  estimate, we can take the eikonal limit where the B-exchange as a closed string is exchanged between two open string dipoles. As a result, the  coupling is of order $g_{B\psi}\sim \sqrt{g_s}$, with the string coupling $g_s=\lambda/4\pi N_c$ 
(pure AdS geometry). For $\lambda\sim 10$, we obtain $g_{B\psi}\sim 0.5$.

Alternatively, if we  use the identification (\ref{BOUNDARYOP}) for the boundary operator, then the near forward C-odd gluonic matrix element in the QCD instanton vacuum is about~\cite{Liu:2024}
\bea
\label{BOUNDARYOPX}
&&\langle P' S| d^{abc}G^{a\alpha\beta} G^b_{\alpha\beta}G^{c\mu\nu}
|P S\rangle\nonumber\\
&&\sim \frac{\kappa^2_{I+\bar I}}{\rho^3}
\,f(q\rho) \,
\langle P' S|\overline{\psi}\sigma^{\mu\nu}\psi|P S\rangle
\eea
Here $f(q\rho)$ is the induced 
form factor by an instanton of size $\rho$. For the kinematical range of interest in Fig~\ref{fig:tWplot}, $\rho \sqrt{|t_{\rm min}|}\sim 1$, and $f(q_{\rm min}\rho)\sim 1$. Indeed, 
for a "dense instanton ensemble"~\cite{Shuryak:2021fsu}, the instanton packing fraction is $\kappa_{I+\bar I}\sim 0.7$ with
a mean  instanton size $\rho\sim \frac 13\,{\rm fm}$. It follows that a simple estimate for the 
dual coupling is $g_{B\psi}\sim \kappa^2_{I+\bar I}\sim 0.5$,
in agreement with the string estimate. 
For a "dilute instanton ensemble"~\cite{Schafer:1996wv}, the packing fraction is 
$\kappa_{I+\bar I}\sim 0.1$, with a weaker dual coupling  $g_{B\psi}\sim \kappa_{I+\bar I}^2\sim 0.01$.
The suppression of the gluons compared to the quarks in a topologically active vacuum, at the resolution $1/\rho$, is similar to the
suppression factor noted for the gluons in comparison to the quarks in the nucleon spin budget~\cite{Zahed:2022wae}.

It would be very useful
to carry a lattice simulation check for the QCD instanton vacuum estimate  (\ref{BOUNDARYOPX}).

\begin{figure*}
\hfill
\subfloat[\label{fig:DIFF403a}]{%
     \includegraphics[height=5cm,width=8cm]{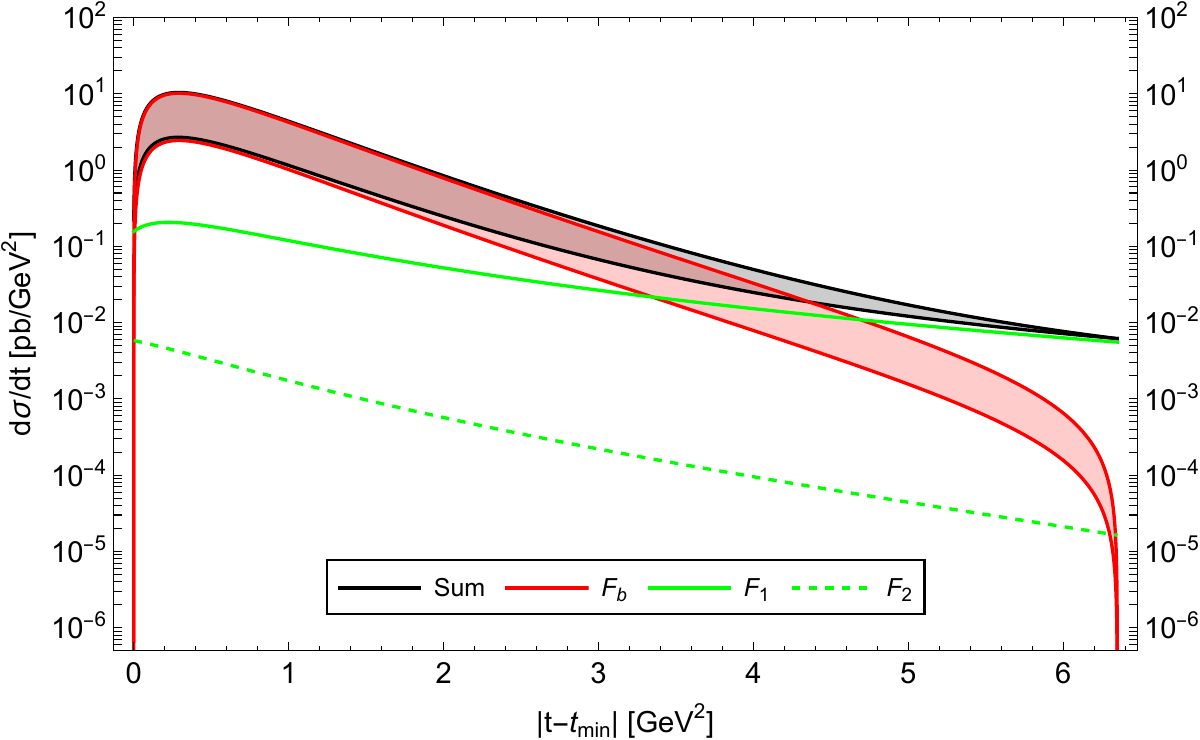}
}\hfill
\subfloat[\label{fig:DIFF403b}]{%
    \includegraphics[height=5cm,width=8cm]{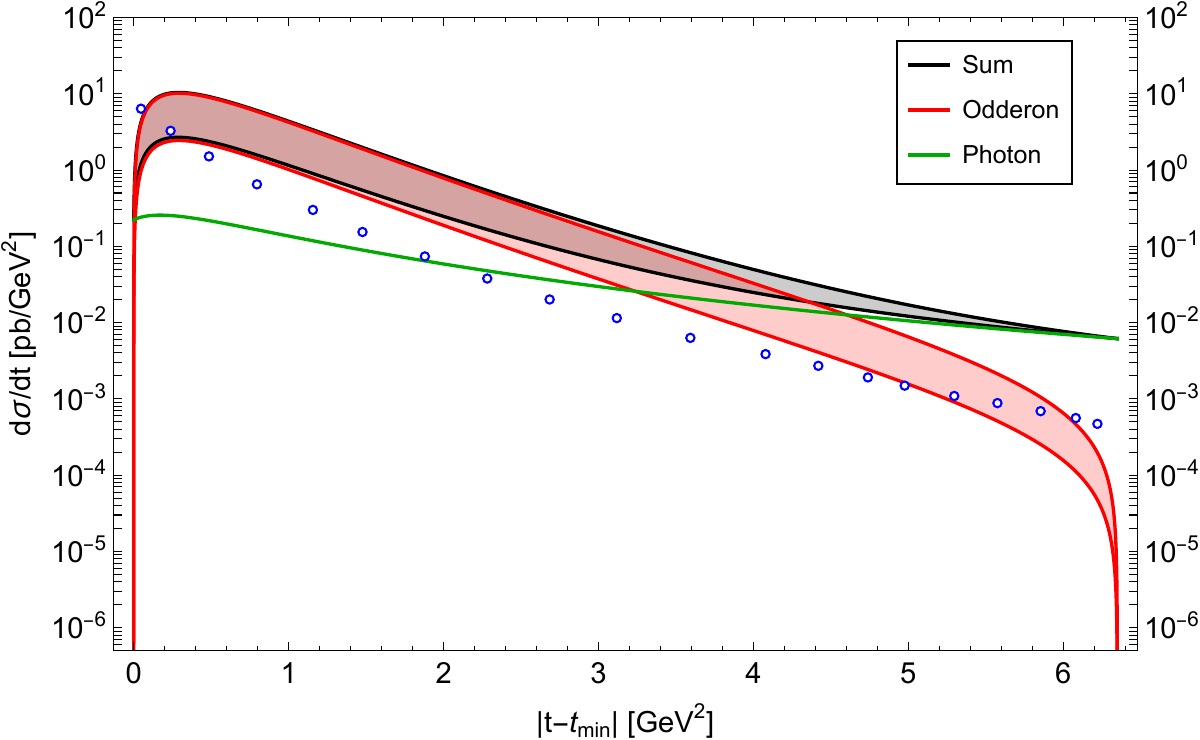}
}
\caption{ 
a: Holographic differential cross section for threshold 
photoproduction of $\eta_c$ at $W=4.3$ GeV,
with the P-wave photon exchange (solid-green: Dirac and dotted-green: Pauli), the Odderon exchange (solid-red with $g_{B\psi}=\{1,0.5\}$) and their coherent sum (solid-black with $g_{B\psi}=\{1,0.5\}$);
b: Holographic differential cross section as in a, compared to
the Primakoff photon exchange estimate (open-blue-circles) in~\cite{Jia:2022oyl}.}
\end{figure*}
\begin{figure*}
\hfill
\subfloat[\label{fig:differential10}]{%
    \includegraphics[height=5cm,width=8cm]{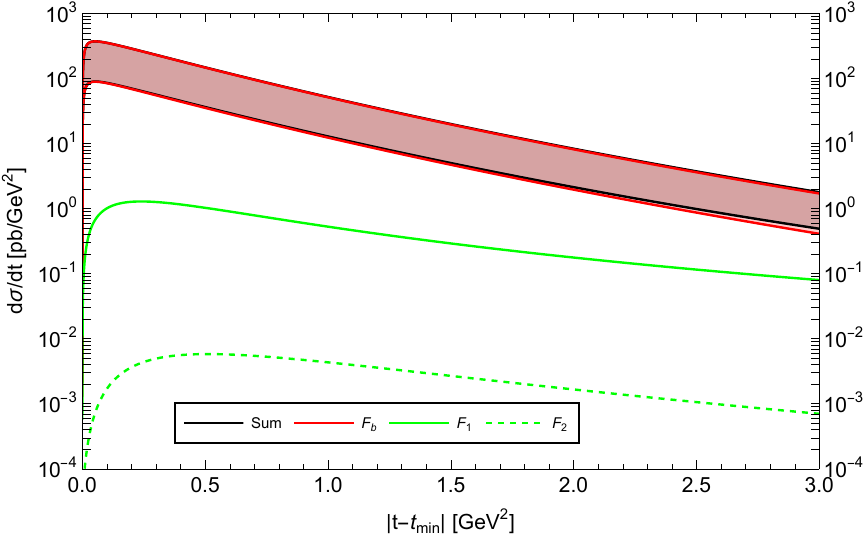}
}\hfill
\subfloat[\label{fig:DIFF4}]{%
  \includegraphics[height=5cm,width=8cm]{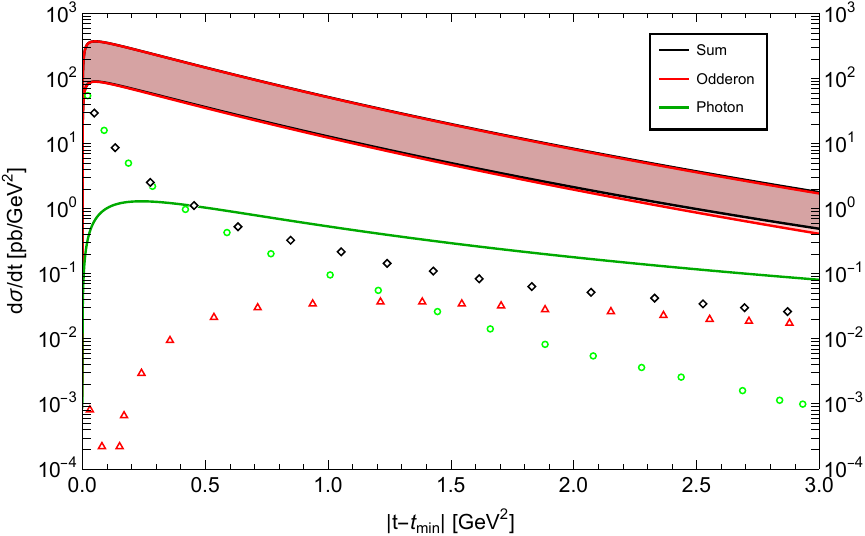}
}
\caption{ 
a: Holographic differential cross section for threshold 
photoproduction of $\eta_c$ at $W=10$ GeV,
with the P-wave photon exchange (solid-green: Dirac and dotted-gree: Pauli), the Odderon exchange (solid-red with $g_{B\psi}=\{1,0.5\}$) and their coherent sum (solid-black with $g_{B\psi}=\{1,0.5\}$);
b: Holographic differential cross section as in a, compared to
the eikonalized dipole approximation in~\cite{Dumitru:2019qec}
with the Primakoff photon exchange (green-diamonds), the Odderon exchange (red-triangles) and their sum (black-diamonds).}
\end{figure*}
\begin{figure*}
\hfill
\subfloat[\label{fig:differential50sep}]{%
    \includegraphics[height=5cm,width=8cm]{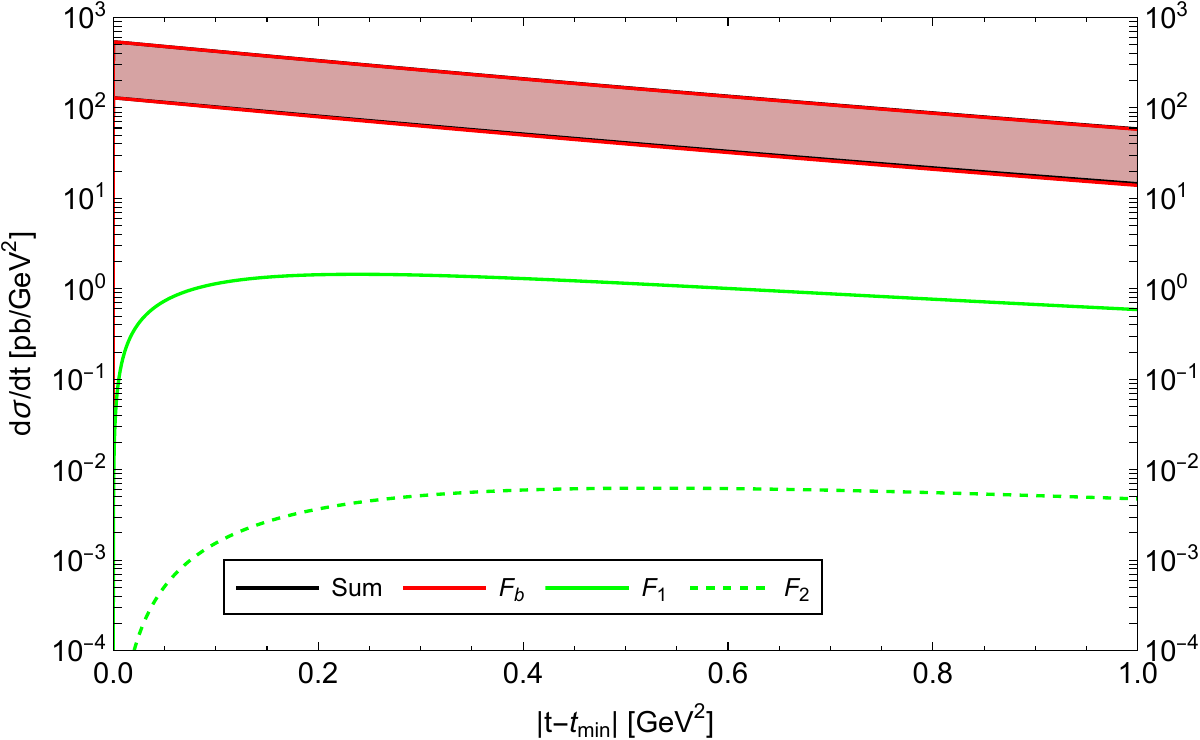}
}\hfill
\subfloat[\label{fig:differential50}]{%
  \includegraphics[height=5cm,width=8cm]{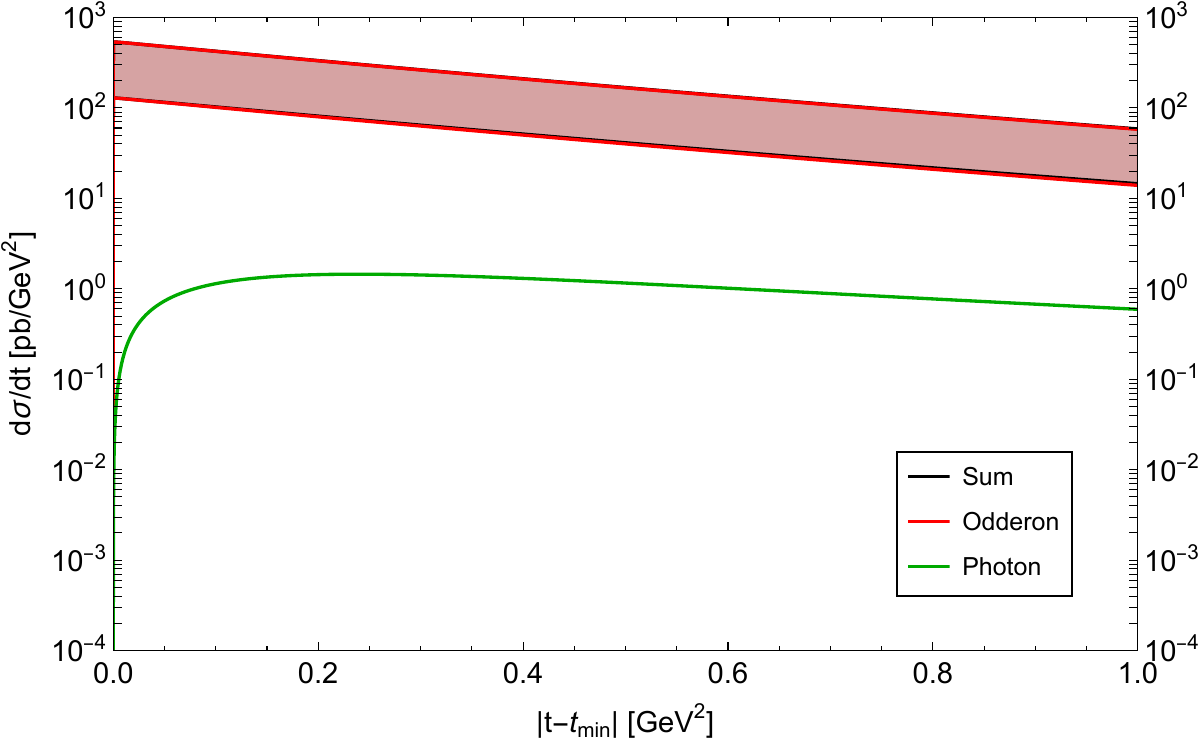}
}
\caption{ 
a: Holographic differential cross section for threshold 
photoproduction of $\eta_c$ at $W=50$ GeV,
with the P-wave photon exchange (solid-green: Dirac and dotted-gree: Pauli), the Odderon exchange (solid-red  with $g_{B\psi}=\{1,0.5\}$) and their coherent sum (solid-black with $g_{B\psi}=\{1,0.5\}$);
b: Holographic differential cross section as in a, but with the photon contribution summed.}
\end{figure*}

\begin{figure*}
\hfill
\subfloat[\label{fig:differential300}]{%
    \includegraphics[height=5cm,width=8cm]{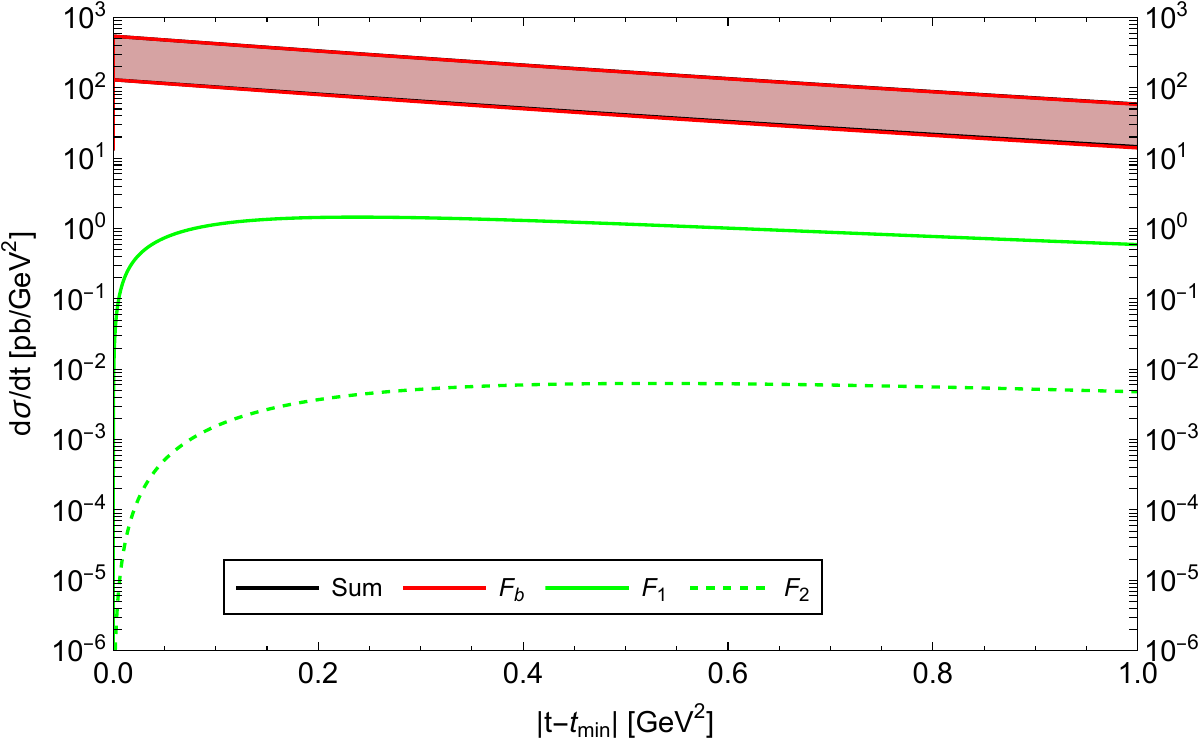}
}\hfill
\subfloat[\label{fig:DDIFFpQCD}]{%
   \includegraphics[height=5cm,width=8cm]{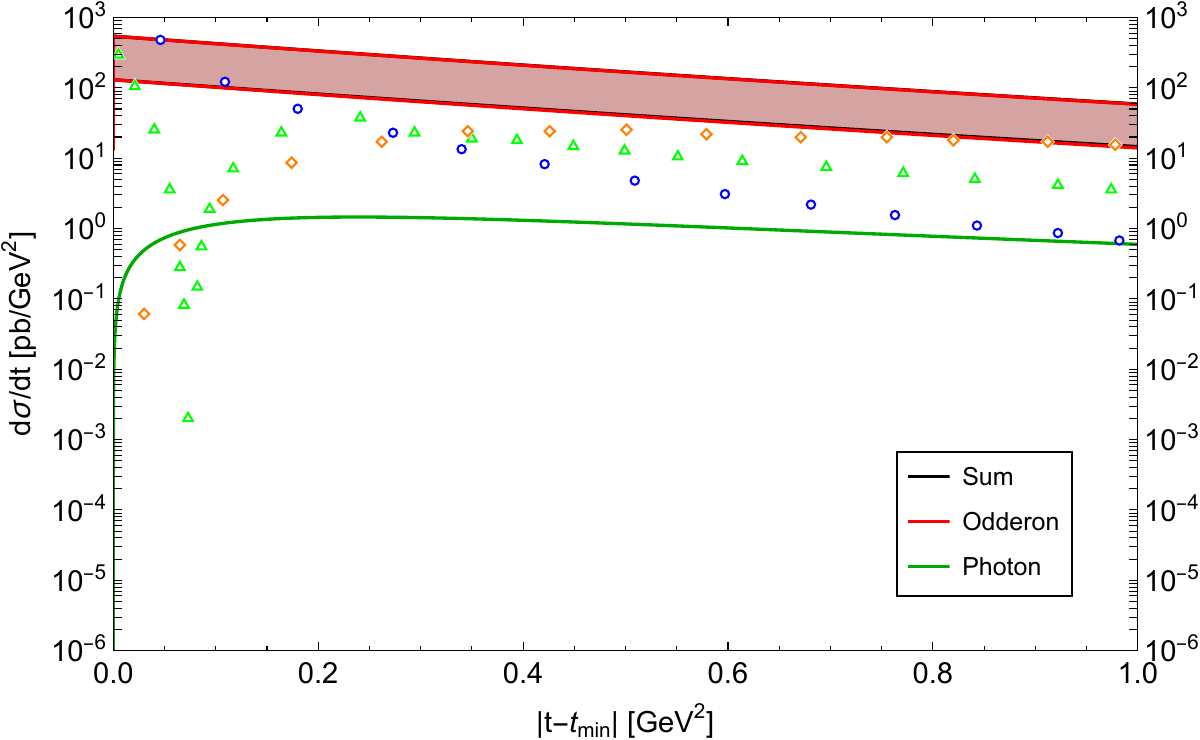}
}
\caption{ 
a: Holographic differential cross section for threshold 
photoproduction of $\eta_c$ at $W=300$ GeV,
with the P-wave photon exchange (solid-green: Dirac and dotted-green: Pauli), the Odderon exchange (solid-red with $g_{B\psi}=\{1,0.5\}$) and their coherent sum (solid-black with $g_{B\psi}=\{1,0.5\}$);
b: Holographic differential cross section as in a, compared to
the Primakoff photon-exchange (blue-open-circles) from~\cite{Jia:2022oyl}, the Odderon model (orange-diamonds) from~\cite{Czy_ewski_1997} and the Odderon model (green-triangles) from~\cite{Bartels_2001}.
}
\end{figure*}

 \begin{figure*}
\hfill
\subfloat[\label{fig:differential11bsep}]{%
    \includegraphics[height=5cm,width=8cm]{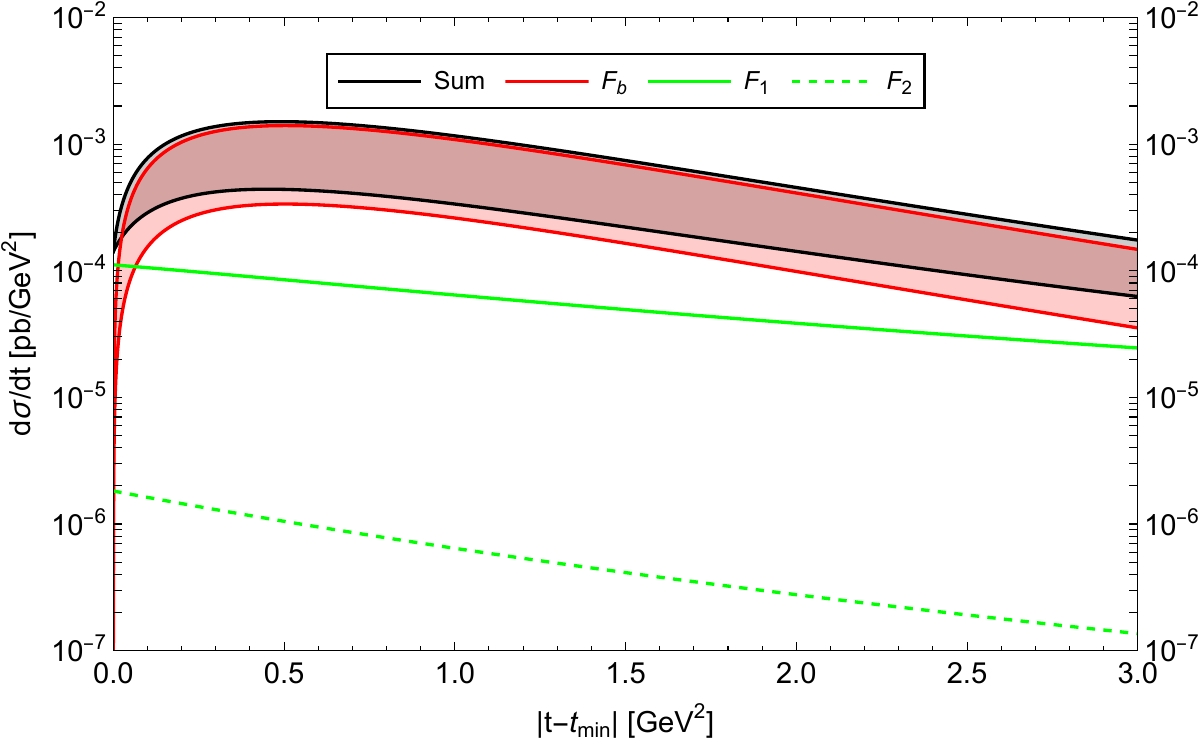}
}\hfill
\subfloat[\label{fig:differential11b}]{%
   \includegraphics[height=5cm,width=8cm]{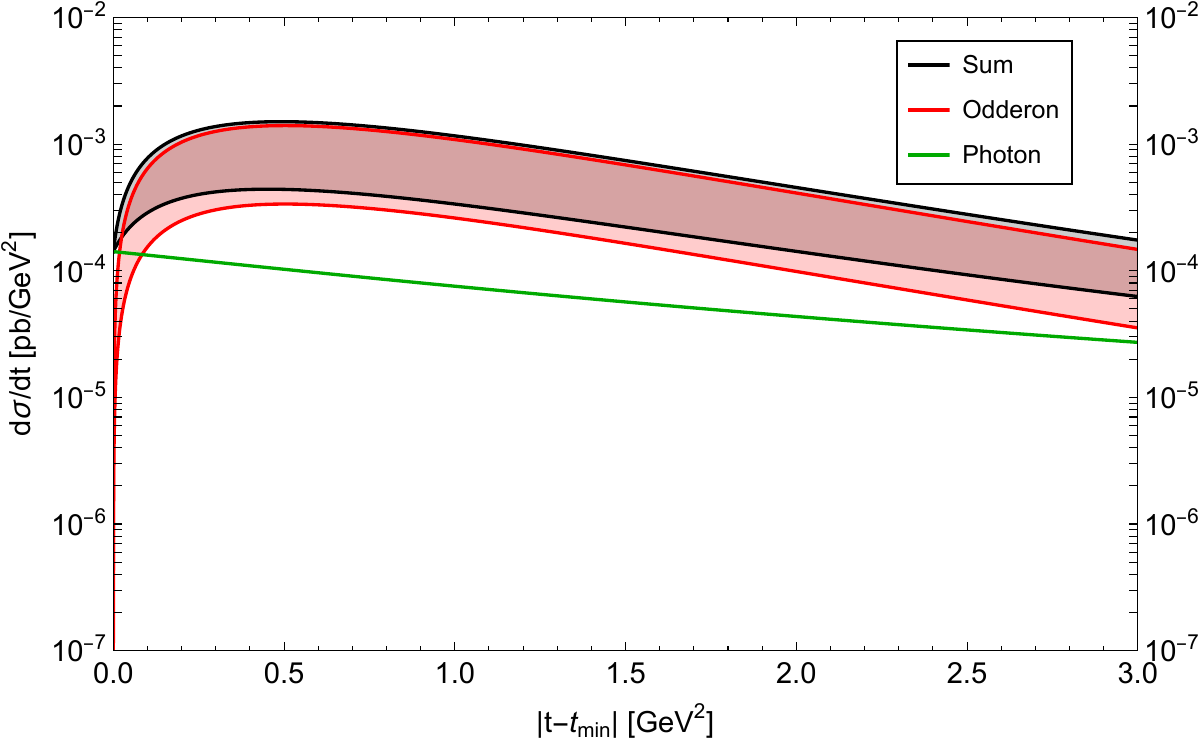}
}
\caption{ 
a: Holographic differential cross section for threshold 
photoproduction of $\eta_b$ at $W=11$ GeV,
with the P-wave photon exchange (solid-green: Dirac and dotted-green: Pauli), the Odderon exchange (solid-red with $g_{B\psi}=\{1,0.5\}$) and their coherent sum (solid-black with $g_{B\psi}=\{1,0.5\}$);
b: Holographic differential cross section as in a, for the 
unseparated contributions.}
\end{figure*}   

\begin{figure*}
\hfill
\subfloat[\label{fig:differentialJpsi4.58}]{%
    \includegraphics[height=5cm,width=8cm]{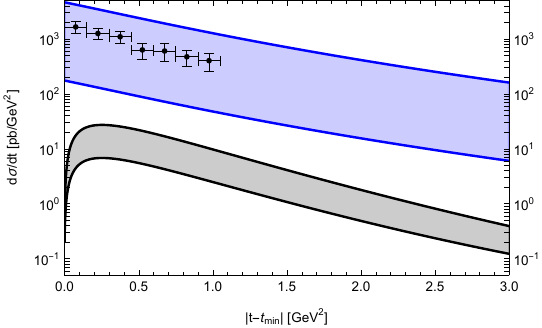}
}\hfill
\subfloat[\label{fig:differentialJpsi4.3}]{%
   \includegraphics[height=5cm,width=8cm]{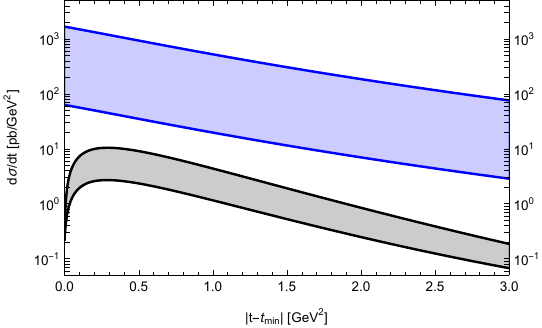}
}
\hfill
\subfloat[\label{fig:differentialJpsi10}]{%
   \includegraphics[height=5cm,width=8cm]{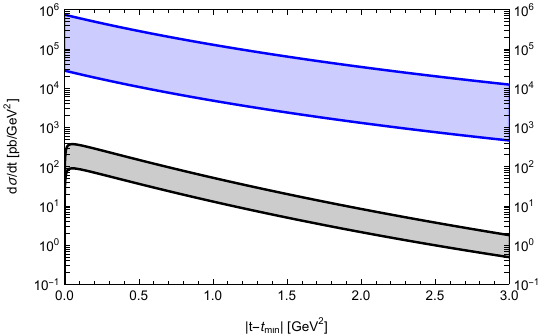}
}\hfill
\subfloat[\label{fig:differentialJpsi50}]{%
   \includegraphics[height=5cm,width=8cm]{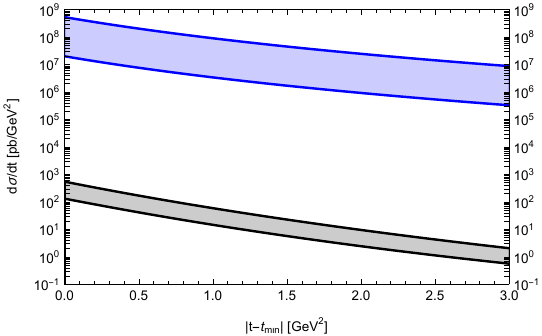}
}\hfill
\subfloat[\label{fig:differentialJpsi300}]{%
   \includegraphics[height=5cm,width=8cm]{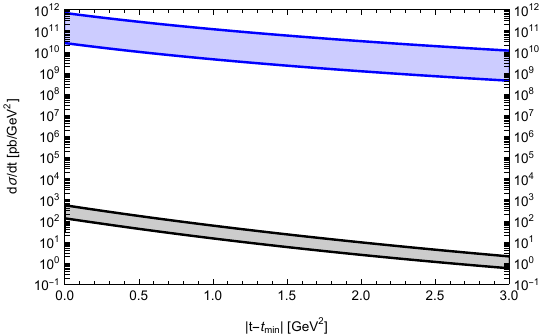}
}\hfill
\subfloat[\label{fig:sigmaJpsi}]{%
   \includegraphics[height=5cm,width=8cm]{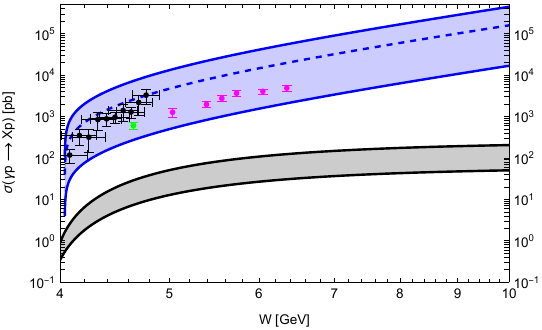}
}
\caption{ 
Holographic differential and total cross sections for threshold 
photoproduction of $J/\Psi$ (blue-shaded) from \cite{Mamo:2019mka}, and the present results for $\eta_c$ (black-shaded with $g_{B\psi}=\{1,0.5\}$): (a) $W=4.58$ GeV, (b) $W=4.30$ GeV, (c) $W=10$ GeV, (d) $W=50$ GeV, (e) $W=300$ GeV; and holographic total cross section for threshold 
photoproduction of $J/\Psi$ (f). The black data points are from GlueX \cite{Ali:2019lzf}. The magenta and green data points are from SLAC \cite{Camerini:1975cy} and Cornell \cite{Gittelman:1975ix}, respectively. }
\label{fig:HOLOXXX}
\end{figure*}

\subsection{Numerical results}
With this in mind, the total cross section for threshold production of $\eta_c$ with $\kappa^\prime$s fixed to the mass spectra, and $g_{B\psi}=\{1,0.5\}$ is 
$\sigma(W=4.3\ \rm GeV)=\{10.3,\ 2.76\}$ pb. It is sensitive to the overall value of $g_{B\psi}$ which we estimated above. In pQCD it corresponds to the fraction of gluons contributing to the nucleon tensor charge as measured by the quarks in (\ref{TENSORCHARGE}). In the numerical results to follow,
all the holographic results will be quoted for $g_{B\psi}=\{1,0.5\}$.

In Fig.~\ref{fig:DIFF403a} we show 
the differential cross section for $W=4.3\ \rm GeV$ versus the threshold-$t$, with the 
P-wave photon contributions (dotted-green: Pauli and solid-green: Dirac), the Odderon contribution (solid-red) and the sum total 
(solid-black). At this center of mass energy
and modulo the  value of $g_{B\psi}$, the differential cross section is dominated by the Odderon exchange near threshold, but is rapidly overtaken by the P-wave photon exchange.  In Fig.~\ref{fig:DIFF403b} we 
compare our results for the differential cross section, to the 
recent estimate using the Primakoff photon exchange estimate (open-blue-dots) in~\cite{Jia:2022oyl}. The holographic result is substantially larger.

In Fig.~\ref{fig:differential10} we show the same differential cross section for $\eta_c$ production at the center of mass energy  $W=10$ GeV. The total cross section at this energy is $\sigma(W=10\ \rm GeV)=\{202,\ 50\}$ pb. Again, the P-wave photon contributions (dotted-green: Pauli and solid-green: Dirac) are compared to the Odderon contribution (solid-red) and the sum total (solid-black). At this energy, the Odderon contribution is dominant throughtout the threshold region. In Fig.~\ref{fig:differential10} the holographic results are compared to the results obtained using the eikonalized dipole approximation for the Odderon in~\cite{Dumitru:2019qec}. The holographic results for the P-wave photon exchange (solid-green), the Odderon (solid-red) and total (solid-black), are compared to Odderon (red-triangle), photon (green-diamond) and total (black-diamond) in~\cite{Dumitru:2019qec}. The sum total of the differential cross section are about comparable at $t=t_{min}$ with the eikonalized results falling off much faster, although there is a substantial difference in the respective contributions, with no crossing in the holographic case in this kinematical range. The difference in the photon contribution stems from the VMD nature of the holographic photon exchange in the bulk in comparison to the simple Primakoff exchange used in~\cite{Dumitru:2019qec} which dwarfs the Odderon contribution at threshold. In Fig.~\ref{fig:differential50} we show the differential cross section at $W=50$ GeV, the relevant kinematical range for the future electron-ion-collider (EIC). The integrated cross sections is given by $\sigma(W=50\ \rm GeV)=\{242,\ 59\}$ pb.

At much higher center of mass energy, say $W=300$ GeV, the integration interval becomes very large since $t_{min}\sim 0$ and $-t_{max}\sim W^2$ and the integrated cross section starts to diverge. Although the Reggeization may start to be important in this kinematical range, we show our un-Reggeized C-odd bulk Odderon exchange in Fig.~\ref{fig:differential300} with the P-wave photon exchange (solid-green: Dirac and dotted-green: Pauli), the Odderon exchange (solid-red) and the sum total (solid-black). In Fig.~\ref{fig:DDIFFpQCD} the holographic results are compared to  the  the estimate for photon-exchange (open-blue-dots) in~\cite{Jia:2022oyl}, and the Odderon model exchange (green-triangle) from~\cite{Bartels_2001} and the Odderon model exchange (orange-diamond) from~\cite{Czy_ewski_1997}. 

In Fig.~\ref{fig:differential11bsep} we show the same differential cross section for photoproduction of $\eta_b$ at $W=11$ GeV, with the P-wave photon exchange (solid-green: Dirac and dashed-green: Pauli), the Odderon exchange (solid-red) and the the sum total (solid-black). The unseparated contributions are shown in Fig.~\ref{fig:differential11b} for the photon (solid-green), Odderon (solid-red) and sum (solid-black). For $\eta_b$ production the integrated cross sections are $\sigma(W=11\ \rm GeV)=\{0.002,\ 0.001\}$ pb and $\sigma(W=22\ \rm GeV)=\{1.20,\ 0.29\}$ pb 
For $\eta_b$, the photon contribution crosses the Odderon contribution twice in the threshold region, underlying the sensitivity to the unfixed overall $g_{B\psi}$ parameter.

In Fig.~\ref{fig:HOLOXXX} we compare the holographic results for the differential  cross section for photoproduction  of $J/\Psi$ (blue-spread)~\cite{Mamo:2019mka}, with  that of $\eta_c$ (black-spread) from this work: (a) $W=4.58$ GeV, (b) $W=4.30$ GeV, (c) $W=10$ GeV, (d) $W=50$ GeV, (e) $W=300$ GeV. The   holographic total cross section for threshold  photoproduction of $J/\Psi$ and its comparison to $\eta_c$ is shown in (f). The black data points are from GlueX \cite{Ali:2019lzf}. The magenta and green data points are from SLAC \cite{Camerini:1975cy} and Cornell \cite{Gittelman:1975ix}, respectively. The holographic result from \cite{Mamo:2019mka} uses  the normalization constant $\mathcal{N}=4.637\pm3.131$, and replaces $A(t)$ by $A(t)+\eta^2\,D(t)$ in Eq.VIII.58 of \cite{Mamo:2019mka}, where $\eta$ is the skewness parameter given explicitly by Eq.III.33 in~\cite{Mamo:2022eui}. We have used the holographic gluonic gravitational form factors $A(t)$ and $D(t)=4C(t)$ extracted by the $J/\Psi-007$ collaboration at JLab \cite{Duran:2022xag}. Note that we have ignored the D-term for the total cross section (f). Also note that the upper limit in the shaded region corresponds to $\mathcal{N}=4.637+3.131=7.768$ used by the $J/\Psi-007$ collaboration at JLab \cite{Duran:2022xag}.

\section{Conclusions}
\label{SEC5}
We have analyzed the differential and integrated cross sections for photoproduction of heavy pseudoscalars $\eta_{c,b}$ in the threshold regime using dual gravity. In this limit, we have suggested that the dominant contribution stems from the exchange of a Kalb-Ramond $B_2$-field in the bulk, which is the dual of 
a $1^{+-}$ glueball.  The glueballs are sourced by a twist-5 boundary operator, which we have argued to be tied to the tensor coupling of Dirac fermions in the bulk as dual to nucleons, modulo an overall constant $g_{B\psi}$ not fixed by holography. This gluon mediated constant was estimated to be small, using the QCD instanton vacuum at low resolution.

A possible measure of the diffractive gluon mediated photoproduction
of heavy mesons near threshold would bring an important insight on the 
C-odd gluonic mass content of the proton, at about the nucleon mass resolution. It would also be an important precursor for the elusive
Odderon, expected to set in at higher energies through Reggeization 
of the C-odd glueballs. 

Near threshold, the $\eta_{c,b}$ photoproduction cross sections through diffractive $1^{+-}$ glueballs are shown to be very sensitive to the value of this coupling $g_{B\Psi}$. This notwithstanding, we have found
that the ensuing diffractive differential cross sections overtake the P-wave photon mediated differential cross sections for $g_{B\Psi}=0.5-1$, as suggested by both a string estimate, 
and an estimate using the (dense) QCD instanton vacuum for the dual boundary operator. The production is depleted by almost two orders of magnitude, say  for  $g_{B\psi}=0.01$ using the (dilute) QCD instanton vacuum for the dual boundary operator. These observations hold for
the current electron facility at JLab with $W\sim 5\, {\rm GeV}$, and the future electron facility at the EIC with $W\sim 50\,{\rm GeV}$.

\begin{acknowledgments}
I.Z. is supported by the Office of Science, U.S. Department of Energy under Contract No. DE-FG-88ER40388,  and in part within the framework of the Quark-Gluon Tomography (QGT) Topical Collaboration, under contract no. DE-SC0023646.
F.H. is supported by the Austrian Science Fund FWF, project no. P 33655-N and the FWF doctoral program Particles \& Interactions, project no. W1252-N27. K.M. is supported by U.S. DOE Grant No. DE-FG02-04ER41302.

\end{acknowledgments}

\appendix

\section{Bulk fields}
\label{app:BulkFields}
\subsection{Bulk Dirac fermions}
Throughout we will use the definitions and partial results in~\cite{Mamo:2022jhp}, to which we also refer for further details. The free bulk Dirac action in terms of the nucleon doublet
\be
\Psi_{1,2}\equiv\begin{pmatrix}
    \Psi_{p1,2}\\
    \Psi_{n1,2}
\end{pmatrix}
\ee
is given by
\begin{widetext}
\be
    S_F=\frac{1}{2g_5^2}\int d^5xe^{-\phi(z)}\sqrt{g}\left(\frac{i}{2}\overline{\Psi}_{1,2}e^N_A\Gamma^A\left(\overrightarrow{D}^{L,R}_N-\overleftarrow{D}_N^{L/R}\right)\Psi_{1,2}-(\pm M+V(z))\overline{\Psi}_{1,2}\Psi_{1,2}\right),
    \label{eq:FreeFermionAction}
\ee
with $V(z)=\kappa_N^2z^2,\ \omega_{\mu\nu z}=-1/z\eta_{\mu\nu}$, anomalous dimension $ M=\pm(\Delta-2)=\pm(\tau-3/2)$ and
\bea
\overrightarrow{D}_N^{L,R}&=&\overrightarrow{\partial}_N+\frac{1}{8}\omega_{NAB}\left[\Gamma^A,\Gamma^B\right]-iX^a_NT^a\nonumber\\
\overleftarrow{D}_N^{L,R}&=&\overleftarrow{\partial}_N+\frac{1}{8}\omega_{NAB}\left[\Gamma^A,\Gamma^B\right]+iX^a_NT^a.\nonumber\\
\eea
The equations of motion governed by \eqref{eq:FreeFermionAction} are given by
\be
\left(ie^N_A\Gamma^A D_N^{L,R}-\frac{i}{2}(\partial_N\phi)e^N_A\Gamma^A-(\pm M+V(z)\right)\Psi_{1,2}=0,
\label{eq:fermionEOM}
\ee
with the normalizable solutions in the bulk
\bea\label{SolutionFermions}
\Psi_1(p,z;n)&=&\psi_R(z;n)\Psi^0_{R}(p)+ \psi_L(z;n)\Psi^0_{L}(p)\nonumber\\
\Psi_2(p,z;n)&=&\psi_R(z;n)\Psi^0_{L}(p)+ \psi_L(z;n)\Psi^0_{R}(p),
\eea
where
\be
&&\psi_R(z;n)=z^{\Delta}\times\tilde{\psi}_R(z;n)=
z^\Delta\times\bigg(n_R \,{{\xi}_N^{\tau-\frac{3}{2}}}
			L^{(\tau-2)}_n({\xi}_N)\bigg)\nonumber\\
&&\psi_L(z;n)=z^{\Delta}\times\tilde{\psi}_L(z;n)=
z^\Delta\times \bigg({n}_L\, {{\xi}_N^{\tau-1}} 
			L^{(\tau-1)}_n({\xi}_N)\bigg),
\ee
\end{widetext}
and $\Psi_{R/L}^0(p)=P_\pm u(p),\ \overline{\Psi}_{R/L}^0(p)=\overline{u}(p)P_\mp$ their respective chiral projections. Here $\Delta=\tau + \frac{1}{2}$, ${\xi}_N={\kappa}_N^2z^2$  and $L_n^{(\alpha)}({\xi_N})$ are the generalized Laguerre polynomials. The free boundary spinors are normalized to
\be
\overline{u}(p)u(p)=2M_N.
\ee
The $\tilde\psi_{L,R}$ are normalized in the bulk
\be
\int dz\, e^{-\kappa_N^2z^2}\,
\frac 1{z^{2\tau-3}}\,
\tilde\psi_{L,R}(z;n)\tilde\psi_{L,R}(z;n')=\delta_{nn'}
\nonumber\\
\ee
with
\bea
n_R&=& n_L \sqrt{\tau-1+n}\nonumber\\
n_L&=&\frac 1{{\kappa_N}^{(\tau-2)}}
\bigg(\frac{2\Gamma(n+1)}{\Gamma(\tau+n)}\bigg)^{\frac 12}.
\eea
The mass spectrum resulting from the non-normalizable modes of \eqref{eq:fermionEOM} displays Regge behaviour
\be
m_n^2=4{\kappa}_N^2(n+\tau-1).
\ee
The bulk-to-boundary Dirac field following from the non-normalizable solutions to the Dirac equation in the bulk \eqref{eq:fermionEOM} are given in terms of Kummer functions
\be
\label{NNZ}
\tilde\psi_R(p, z)=N_R\,U\bigg(-\frac {p^2}{4\kappa_N^2}, 3-\tau, \xi_N\bigg)\nonumber\\
\tilde\psi_L(p, z)=N_L\,U\bigg(-\frac {p^2}{4\kappa_N^2}, 2-\tau, \xi_N\bigg).
\ee
with $N_R/N_L=p/2\kappa_N$ and
$$N_L=\frac{\Gamma\bigg(\tau-1-\frac{p^2}{4\kappa_N^2}\bigg)}{\Gamma(\tau-1)}$$
Note that (\ref{NNZ}) can be recast as the resummed Regge poles
\bea
\tilde\psi_R(p, z)&=&\sum_{n=0}^\infty \frac{f_n p\,\tilde\psi_R(n; z)}{p^2-m_n^2}\nonumber\\
\tilde\psi_L(p, z)&=&\sum_{n=0}^\infty \frac{f_n m_n\,\tilde\psi_L(n; z)}{p^2-m_n^2}
\eea
with the couplings $f_n=2\kappa_N/(n_R\Gamma(\tau-1))$. For later convenience we also define $F_n=-m_n f_n$.


During the reduction of the chiral spinors in the interaction terms to 4D, we will encounter the following expressions
\bea
\overline{\Psi}_1\gamma^\mu\Psi_1 +\overline{\Psi}_2\gamma^\mu\Psi_2&=&(\psi_R^2+\psi_L^2)\overline{u}\gamma^\mu u\nonumber\\
\overline{\Psi}_1\gamma^\mu\gamma^5\Psi_1 -\overline{\Psi}_2\gamma^\mu\gamma^5\Psi_2&=&(\psi_R^2-\psi_L^2)\overline{u}\gamma^\mu u\nonumber\\
\overline{\Psi}_1\gamma^\mu\gamma^\nu\Psi_1 -\overline{\Psi}_2\gamma^\mu\gamma^\nu\Psi_2&=&2\psi_R\psi_L\overline{u}\gamma^\mu\gamma^\nu u.\nonumber\\
\eea

\subsection{Bulk pseudoscalar fields}
The pseudoscalar fluctuations are contained in $A_z$ of the 5d vector field $A_M$. The equation of motion following from the quadratic part of the action \eqref{XACTIONX} is given by
\be
\partial_M\left(\sqrt{g}e^{-\phi}F^{MN}\right)=0
\ee
In particular we obtain
\bea
\Box V^\mu+ze^\phi\partial_z\left(e^{-\phi}\frac{1}{z}\partial_z V^\mu\right)&=&0\nonumber\\
\Box V_z-\partial_z\left(\partial_\mu V^\mu\right)=0\nonumber\\
\eea
subject to the gauge condition
\be
\partial_\mu V^\mu + ze^\phi\partial_z\left(e^{-\phi}\frac{1}{z} V_z\right)=0
\ee
The normalizable modes are given by
\be
\phi_n(z)=c_n \kappa z L_n(\kappa^2z^2)
\ee
with the normalization fixed by
\be
\int\sqrt{g}e^{-\phi}e^{-4A(z)}\phi_m(z)\phi_n(z)=\delta_{mn}
\ee
and $c_n=\frac 12$. The ensuing mass spectrum follows as
\be
m_n^2=4\kappa^2(n+1),
\ee
and displays again the expected Regge behaviour. Note that the decay constant given by
\be
F_n=\frac{1}{g_5}\left(e^{-\phi}\frac{1}{z'}\partial_{z'}\phi_n(z')\right)\bigg|_{z'=0}
\ee
is strictly divergent at $z'=0$. The correct 
UV boundary condition should be set by a heavy brane with $z'\sim \frac 1{m_c}$.
With this in mind the bulk wavefunctions can be written as
\be
\phi_n(z)=-\frac{f_n}{m_n}\times 4g_5(n+1)\kappa z L_n(\kappa^2z^2),
\ee
with $f_n=-F_n/m_n$ fixed to its experimental value in the main text. 

\subsection{Bulk spin-1 fields}
\subsubsection{Top-down Kalb-Ramond field}
In type-II SUGRA the fields $B_2$ and $C_2$ (IIA) and $C_3$ (IIB) are mixed via a topological mass term. In particular a consistent solution to the equations of motion studied in \cite{Brower:2000rp} is only given by $B_{\mu\nu}$ and $C_{\mu\tau r}$ for $1^{+-}$ and $B_{\mu z}$ and $C_{\mu\nu\tau}$ for $1^{--}$, where $\tau$ is the supersymmetry breaking compactified direction of the Witten model \cite{Witten:1998zw}. For example, the relevant linearized type IIA equations of motion in 10D string frame are given by
\begin{widetext}
 \begin{equation}
    \begin{split}
        \nabla_O\left(e^{-2\phi}H^{OMN}\right)-\frac{1}{2!\cdot(4!)^2\sqrt{-g}}\epsilon^{MNO_1\hdots O_8}F_{O_1\hdots O_4}F_{O_5\hdots O_8}&=0,\\
        \nabla_P F^{PMNO}-\frac{1}{3!\cdot 4!\sqrt{-g}}\epsilon^{MNOP_1\hdots P_7}H_{P_1P_2P_3}F_{P_4\hdots P_7}&=0,\\
    \end{split}
\end{equation}   
\end{widetext}
which are coupled through a non-vanishing $F_4$ flux generated by the $N_c$ color branes. To solve them for the $1^{--}$ polarization, we start with the radial ansatz 
\begin{equation}
    C_{\mu\nu\tau}=\frac{a(r)}{g_s}\tilde{C}_{\mu\nu}(x^\mu),\quad B_{\mu r}=b(r)\eta_{\mu\kappa}\epsilon^{\kappa\nu\rho\sigma}\partial_\nu \tilde{C}_{\rho\sigma}(x^\mu),
\end{equation}
where we suppress the plane wave factors $e^{ikx}$. The equation of motion for $H_3$ gives
\be
 b(r)=\frac{3}{2\Box}e^{4\lambda(r)}a(r)
\ee
and upon substituting this result into the equation of motion for $C_3$ we get the equation of motion for the $1^{--}$ glueball. The factor $e^{4\lambda}$ pertains to the metric factors on $\mathcal{M}_4$ with $ds^2_{\mathcal{M}_4}=e^{2\lambda(r)}\eta^{\mu\nu}dx^\mu dx^\nu$ and $r$ is the holographic coordinate. One can check that all other linearized equations of motion resulting from the type IIA closed string action are satisfied, and the Lagrangian is diagonal. To project out the three polarizations of a massive spin-1 field, we use $\tilde{C}_{\rho\sigma}(x^\mu)=\frac{1}{\sqrt{\Box}}\epsilon_{\rho\sigma}^{\ \ \kappa\lambda}\partial_\kappa V_\lambda(x^\mu)$. Which ultimately leads us to
\bea
C_{\mu\nu\tau}=\frac{2a(r)}{\sqrt{\Box}g_s}\star F_{\mu\nu},\ B_{\mu r}=\frac{3}{\sqrt{\Box}}e^{4\lambda}a(r)V_\mu,\nonumber\\
\eea
and a canonically normalized kinetic term pertinent for a spin-1 field. This projection will also result in the correct kinetic term in \eqref{XACTIONX}. For the $1^{+-}$ polarization the situation is precisely reversed: $C_{\mu\tau r}\sim V_\mu$ and $B_{\mu\nu}\sim F_{\mu\nu}$. Note that the relevant interactions originate from the Chern-Simons term, which leads to the correct parity assignments of the interactions.
\subsubsection{Soft-wall}
As discussed above, after dimensional reduction, the  SUGRA action for the Kalb-Ramond field reduces to an effective spin-1 action in 5d. The subsequent formulas thus als hold for the bulk photon fields, with mass scale set by $\kappa_\gamma$ and the corresponding bulk wave functions substituted by $J_b(m_n,z)\to J(m_n,z)$ and $\mathcal{V}_b(K,z)\to \mathcal{V}(K,z)$. Following \cite{Grigoryan:2007my} we describe the t-channel exchange of the Kalb-Ramond field via the exchange of a massive spin-1 field with bulk wave function
\be
\phi_n(z)=c_n\kappa_b^2z^2L_n^1(\kappa_b^2z^2)\equiv J_b(m_n,z)
\ee
Note that the coupling of the closed string sector is twice that of the open string sector, hence the dilaton is given by $\phi=\kappa_b^2z^2=(2\kappa_\gamma)^2z^2=(2\kappa)^2z^2$ and $g_5$ is to be understood as $\tilde{g_5}$. The wavefunctions are normalized via
\be
\int dz \sqrt{g}e^{-\phi}e^{-4A(z)}\phi_m(z)\phi_n(z)=\delta_{mn},
\ee
giving $c_n=\sqrt{2/(n+1)}$. With a decay constant given by
\be
F_n=\frac{1}{g_5}\left(e^{-\phi}\frac{1}{z'}\partial_{z'}\phi_n(z')\right)\bigg|_{z'=0}=-\frac{2}{g_5}c_n(n+1)\kappa_b^2,\nonumber\\
\ee
and a Reggeized mass spectrum
\be
m_n^2=4\kappa^2(n+1).
\ee
Fixing the mass spectrum to correspond to the lowest glueball state on the Odderon trajectory, we would obtain
\be
\kappa_b=1.925{\ \rm GeV}^{-1},\ \kappa_b=1.47{\ \rm GeV}^{-1},
\ee
for the $1^{--}$ glueball with mass $M=3.85$ GeV and the $1^{+-}$ glueball with mass $M=2.94$ GeV, respectively.
For our computations, we fix the mass spectrum by the rho meson pole of the time-like photon bulk-to-bulk propagato. With a rho mass of $m_\rho=0.775$ GeV we thus obtain 
\bea
\kappa_\gamma=0.3875\ \rm GeV^{-1}.\nonumber\\
\kappa_b=2\kappa_\gamma=0.775\rm GeV^{-1}.
\eea
With this in mind, we can rewrite the bulk wavefunction as
\be
\phi_n(z)=\frac{f_n}{m_n}\times 2g_5 \kappa_b^2z^2L_n^1(\kappa_b^2z^2)
\ee
where $f_n=-F_n/m_n$. At the production threshold, the external wavefunctions are localized at the boundary. In this limit, the bulk-to-bulk propagator in the mode sum representation
\be
G_1(z,z')=\sum_n\frac{\phi_n(z)\phi_n(z')}{k^2-m_n^2}
\ee
reduces to
\bea
G_1(z\to 0,z')&\approx& \frac{\phi_n(z\to 0)}{-g_5 F_n}\sum_n\frac{-g_5 F_n\phi_n(z')}{k^2-m_n^2}\nonumber\\
&=&\frac{z^2}{2}\times V(k,z').\nonumber\\
\eea
For spacelike $k^2=-K^2$ we thus have
\be
G_1(z\to 0,z')\approx\frac{z^2}{2}\sum_n\frac{g_5 F_n\phi_n(z')}{K^2+m_n^2}=\frac{z^2}{2}\times \mathcal{V}(K,z'),\nonumber\\
\label{eq:b2bSpacelike}
\ee
with
\bea
\mathcal{V}(K,z)&=&\kappa_b^2z^2\Gamma(1+a_K)\mathcal{U}(1+a_K,2,\kappa_b^2z^2 )\nonumber\\
&=&\kappa_b^2z^2\int_0^1\frac{dx}{(1-x)^2}x^{a_K}\exp\left[-\frac{x}{1-x}\kappa_b^2z^2\right],\nonumber\\
\eea
and with $a_K=K^2/4\kappa_b^2$, $\mathcal{U}(a,b,c)$ the confluent hypergeometric function of the second kind and normalized to $\mathcal{V}(0,z)=\mathcal{V}(K,0)=1$. Note that the bulk-to-boundary propagator for an on-shell photon is trivially represented by $\mathcal{V}(0,z)=1$ in the Witten diagram of Fig.~\ref{fig:hOdderonEx}.

\subsubsection{Hard-wall}
In the hard-wall model ($\kappa=0$), the normalizable bulk wave functions are given by
\be
\phi_n(z)=c_n z J_1(m_n,z),
\ee
where $c_n=\sqrt{2}/z_0 J_1(m_nz_0)$ and the mass spectrum is fixed by the n-th root, $r_n$, of the Bessel function
\be
J_0(m_nz_0)=0.
\label{eq:HWmassSpectrum1}
\ee
Fixing the lowest mass to the $\rho$ meson pole in the photon bulk-to-bulk propagator we obtain $z_0=3.103\text{ GeV}^{-1}$, which we use as a hard-wall cut-off in the divergent parts of the Chern-Simons interactions.


\section{Couplings in Witten-Sakai-Sugimoto model}
\label{SSTYPEII}
For reference, we also give the type IIA $1^{\pm-}$ vector couplings obtained in the Witten-Sakai-Sugimoto model, which were first studied in \cite{Brunner:2018wbv} and recently extended and completed in \cite{Hechenberger:2024piy}. The interactions are fully governed by the Chern-Simons term, which we assumed to also carry over to the soft-wall computations in the main text. For the $1^{--}$ vector fluctuation one obtains
\bea
    \mathcal{L}_{G_V\Pi \cal{V}}
    &=&\frac{1}{M_V}g_1^{\cal{V}} \rm{tr}\,\Pi\, F_{\mu\nu}\star F^V_{\mu\nu}
    \label{eq:LagGVsV}
\eea
where
\bea
        g_1^{\mathcal{V}}&=&\frac{9}{16}\sqrt{\frac{\kappa}{\pi}}\frac{1}{M_{\mathrm{KK}}^2 R^3}\int d z\frac{1}{z}\partial_z(z M_4(z))\nonumber\\
        &=&\frac{0.31}{\sqrt{N_c}}
\eea
$R$ is the AdS radius, $M_4(z)$ the normaliable mode of the bulk spin-1 glueball, $\mathcal{V}(q,z)$ is the photon bulk-to-boundary propagator and the numerical value is obtained for an on-shell photon with $\mathcal{V}(0,z)=1$. Note that $z$ here is related to the radial coordinate in the the Sakai-Sugimoto model by $$1+z^2=\left(\frac{U}{U_{\rm KK}}\right)^3.$$ The mass scale is again fixed by the rho meson pole, which gives $M_{\rm KK}=949$ MeV and the 't Hooft coupling is fixed by the pion decay constant to be $\lambda=16.63,\ \kappa=\lambda N_c/216\pi^3$. The corresponding coupling for the $1^{+-}$ fluctuation is given by
\be
    \mathcal{L}_{G_{PV}\Pi \cal V}
    &=&-\frac{1}{M_{PV}}b_1^{\cal{V}} \rm{tr}\,\Pi\, F_{\mu\nu}F_{\mu\nu}^{PV},
\ee
where
\bea
    b_1^{\mathcal{V}}&=&\frac{45}{8}\sqrt{\frac{\kappa}{\pi}}\frac{1}{M_{\mathrm{KK}}^2R^3}\int\frac{d z}{\sqrt{1+z^2}}N_4(z)\nonumber\\
    &=&\frac{2.25}{\sqrt{N_c}},
\eea
with the normalizable $1^{+-}$ bulk mode $N_4(z)$. Taken at face value we have $-b_1^{\mathcal{V}}/2M_{PV}\approx-0.2\ \mathrm{ GeV}^{-1}$ and $\frac{N_c}{24\pi^2}f_{\eta_c}/m_{\eta_c} \mathbb{V}_{B\eta\gamma}\approx-0.05\ \mathrm{GeV}^{-1}$, where we used $M_{PV}=3.2 $ GeV from unquenched lattice QCD \cite{Gregory:2012hu}. However, note that the Witten-Sakai-Sugimoto model treats quarks as massless and quarkonia receive large mass contributions from their quark content.

\section{Kinematics}
\label{app:kin}
\begin{figure*}
\subfloat[\label{fig:etaCtW}]{%
  \includegraphics[height=5cm,width=.45\linewidth]{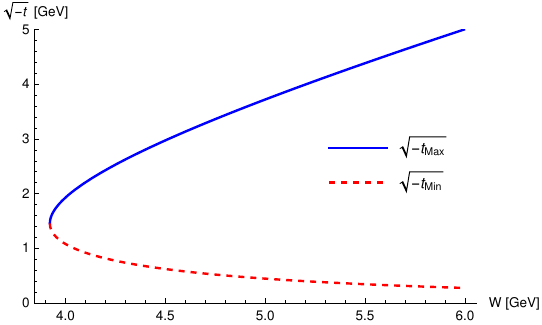}%
}\hfill
\subfloat[\label{fig:etaBtW}]{%
  \includegraphics[height=5cm,width=.45\linewidth]{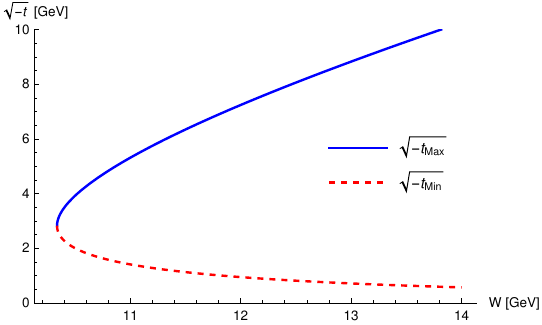}%
}
\caption{Minimal and maximal transverse momentum transfer $t_{\rm min},\ t_{\rm max}$ in the physical region for $\eta_c$ (a) and $\eta_b$ (b) versus $W=\sqrt{s}$. The photon momentum is taken to be at the optical point $q^2=-Q^2=0$ and the hadron masses are given by $M_N=0.938$ GeV, $m_{\eta_c}=2.984$ GeV and $M_{\eta_b}=9.399$ GeV.}
\label{fig:tWplot}
\end{figure*}
The invariants for the meson photoproduction are $s$ and $t$, respectively. Here $s=(q_1+p_1)^2$ is related to the center of mass energy $W=\sqrt{s}$ and $t=\Delta^2$ is related to the momentum transfer $\Delta^\mu=(p_2-p_1)^\mu$. For photoproduction $Q^2=0$, but leptoproduction can also be analyzed with minor variations. In the center of mass frame, the four-momenta of the incoming photon, incoming proton, outgoing proton and outgoing meson $X$ are denoted by $q_1$, $p_1$, $p_2$, and $q_2$ respectively. Each external state is given by the on-shell conditions defined as 
\begin{align*}
p_1^2=p_2^2&=M_N^2\ , & q_1^2=&0\ , & q_2^2=M_X^2.
\end{align*} 
Using the on-shell conditions, the four-momenta in the center of mass frame, can be written as
\begin{align}
\label{eq:kinematics}
q_1=&\left(\frac{s-M^2_N}{2\sqrt{s}},\ 0,\ -\frac{s-M^2_N}{2\sqrt{s}}\right)  & \\[5pt] \nonumber 
q_2=&\left(\frac{s+M_X^2-M^2_N}{2\sqrt{s}},\ -|\vec{p}_{X}|\sin\theta,\ -|\vec{p}_{X}|\cos\theta\right) \\[5pt] \nonumber
p_1=&\left(\frac{s+M^2_N}{2\sqrt{s}},\ 0,\ \frac{s-M^2_N}{2\sqrt{s}}\right)  &  \\[5pt] \nonumber
p_2=&\left(\frac{s-M_X^2+M^2_N}{2\sqrt{s}},|\vec{p}_{X}|\sin\theta,\ |\vec{p}_{X}|\cos\theta \right)
\end{align}
where $M_N$ is the nucleon mass, $M_X$ is the produced meson mass, and $\theta$ is the scattering angle in the center of mass frame. 
The magnitude of the outgoing three-momentum reads
\begin{equation}
|\vec{p}_X|=\left(\frac{[s-(M_X+M_N)^2][s-(M_X-M_N)^2]}{4s}\right)^{1/2}  .  
\end{equation}   
The scattering angle is fixed   by the invariant $t$,
\begin{eqnarray}
    \cos\theta=\frac{2st+(s-M^2_N)^2-M^2_X(s+M_N^2)}{2\sqrt{s}|\vec{p}_X|(s-M_N^2)}\nonumber\\
\end{eqnarray}
with $\bar{p}^\mu=\frac 12(p_1+p_2)^\mu$.
In the threshold limit $\sqrt{s}\rightarrow M_N+M_X$, the momentum transfer $t$ is near the threshold value  $t_{th}$
\begin{equation}
\label{eq:t_th}
    t_{th}=-\frac{M_NM_X^2}{M_N+M_X}
\end{equation}
The kinematically allowed regions are shown on the $(W,\sqrt{-t})$ plane in Fig.~\ref{fig:tWplot} for $\eta_c$ and $\eta_b$, respectively. In the near threshold region $s\gtrsim (M_N+M_X)^2$, the factorization for the proton occurs when the  outgoing meson is heavy enough, so that the proton target moves fast enough to be factorized in partons. In the heavy limit, the incoming and outgoing nucleon velocity is of order 1 modulo $M_N^2/M_X^2$ corrections. In this regime, factorization holds near  threshold for photoproduction, with a non-relativistic outgoing meson with a skewness of order 1~\cite{Ma:2003py,Guo:2021ibg,Sun:2021pyw}.

\section{Top-down fermionic couplings}
Since in type II SUGRA minimal couplings of the form $\sigma^{MN}B_{MN}$ are actually absent due to the strict constraints from supersymmetry, we explore various different top-down couplings in this Appendix. By identifying the fermion couplings by those resulting for the mesinos on the flavor branes we have \cite{Marolf:2003ye}
\begin{widetext}
\bea
S_{Dp}&\supset& i T_{D_p}\int d^{p+1}xe^{-\phi}\sqrt{g}\sum_{1,2}\overline{\Psi}_{1,2}\left(\Gamma^M\breve{D}_M-\breve{\Delta}\right)\Psi_{1,2}\nonumber\\
\label{eq:mesinoAction}  
\eea
with
\bea
\breve{D}_M=D^{(0)(1,2)}+\sigma_1\otimes W_{(1,2)M},\qquad \breve{\Delta}=\Delta^{(1)}+\sigma_1\otimes\Delta^{(2)}
\eea
where
\bea
D^{(0)(1,2)}_M&=&\partial_M+\frac{1}{4}\omega_{MAB}\Gamma^{AB}\pm\frac{1}{4\cdot 2!}H_{MAB}\Gamma^{AB}\nonumber\\
W_{(1,2)M}&=&\frac{1}{8}e^\phi\bigg(\mp F_A\Gamma^A-\frac{1}{3!}F_{ABC}\Gamma^{ABC}\mp\frac{1}{4!}F_{ABCD}\Gamma^{ABCD}\bigg)\Gamma_M\nonumber\\
\Delta_{(1,2)}^{(1)}&=&\frac{1}{2}\left(\Gamma^M\partial_M\phi\pm\frac{1}{2\cdot 3!}H_{ABC}\Gamma^{ABC}\right)\nonumber\\
\Delta_{1,2}^{(2)}&=&\frac{1}{2}e^\phi\left(\pm\frac{1}{2!}F_{A}\Gamma^{A}+\frac{1}{2\cdot 3!}F_{ABC}\Gamma^{ABC}\right),\nonumber\\
\eea
\end{widetext}
and $\Gamma^{A\dots}$ the antisymmetrized product of gamma matrices. By introducing the chiral spin-connection
\be
\omega_M^{(\pm)AB}=\omega_M^{\ AB}\pm \frac{1}{4\cdot 2!}e^{NA} e^{OB} H_{MNO},
\ee
which is amenable to a spin connection with torsion
\be
\tilde{\omega}_M^{\ AB}=\omega_M^{\ AB}+e_N^A e^{BO}\tilde{\Gamma}^N_{MO},
\ee
where $\tilde{\Gamma}$ is the antisymmetric part of the Christoffel symbol, the Kalb-Ramond field can be viewed as a source for torsion, which was first observed in \cite{Scherk:1974mc}.
The couplings arising from $H_3$ in the covariant derivative, and in particular $\Delta^{(1)}$, in \eqref{eq:mesinoAction} for the $1^{+-}$ field $V^\sigma$ in the main text reduce to
\bea
(\psi_L^2+\psi_R^2)\mathcal{V}_b(K,z)\epsilon_{\sigma\alpha\beta\gamma}\sqrt{K^2}V^\sigma\overline{u}(p_2)\gamma^\alpha\gamma^\beta\gamma^\gamma u(p_1),\nonumber\\
\eea
where we note that the $H_{Z\mu\nu}\Gamma^{Z\mu\nu}$ coupling vanishes after the reduction to the 4D spinor is carried out. This means that also the $B_{\mu z}$ fluctuation corresponding to $1^{--}$ does not couple through this term. However the fluctuations $B_{\mu\nu},\ C_{\mu Z}$ and $B_{\mu Z},\ C_{\mu\nu}$ form the physical $1^{\pm -}$ states and we obtain from $F_3\Gamma^{(3)}$ in $\Delta^{(2)}$ for the $1^{--}$ field $V^\sigma$
\bea
2\psi_L\psi_R\mathcal{V}_b(K,z)\epsilon_{\sigma\alpha\beta\gamma}\sqrt{K^2}V^\sigma\overline{u}(p_2)\gamma^\alpha\gamma^\beta\gamma^\gamma u(p_1).\nonumber\\
\eea
Note that both couplings have the correct 5D parity. After the spin sums, the resulting squared matrix elements are highly suppressed at low $K^2$. Other couplings yield  the nucleon axial-tensor charge
\bea
(\overline{u}_L(p_2)-\overline{u}_R(p_2))\sigma^{\mu\nu}u(p_1)=
\overline{u}(p_2)\gamma^5\sigma^{\mu\nu}u(p_1),\nonumber\\
\eea
up to a factor $\sqrt{K^2}$, which is again suppressed in the near-forward regime.

\bibliography{etac}

\end{document}